\documentclass[sigconf, nonacm]{acmart}
\usepackage{algorithm}
\usepackage{algpseudocode}

\usepackage{balance}

\makeatletter
\renewcommand*{\ALG@name}{Listing}

\makeatother

\newcommand{\ScFunction}[2]{%
  \Function{\textsc{#1}}{#2}%
}

\newcommand\vldbdoi{XX.XX/XXX.XX}

\newcommand\vldbavailabilityurl{}
\newcommand\vldbpagestyle{plain} 

\begin{document}
\title{Data Generation for Testing Complex Queries}

\author{Sunanda Somwase}
\affiliation{%
  \institution{Indian Institute of Technology Bombay}
}
\email{sunandasomwase@cse.iitb.ac.in}

\author{Parismita Das}
\affiliation{%
  \institution{Indian Institute of Technology Bombay}
}
\email{parismita@cse.iitb.ac.in}

\author{S. Sudarshan}
\affiliation{%
  \institution{Indian Institute of Technology Bombay}
}
\email{sudarsha@cse.iitb.ac.in}
\begin{abstract}
Generation of sample data for testing SQL queries has been an important task for many years, with applications such as testing of SQL queries used for data analytics and in application software, as well as student SQL queries.  More recently, with the increasing use of text-to-SQL systems, test data is key for the validation of generated queries.
Earlier work for test data generation handled basic single block SQL queries, as well as simple nested SQL queries, but could not handle more complex queries.
In this paper, we present a novel data generation approach that is designed to handle complex queries, and show its effectiveness on queries for which the earlier XData approach is not as effective.  We also show that it can outperform the state-of-the-art VeriEQL system in showing non-equivalence of queries. 
\end{abstract}

\maketitle

\pagestyle{\vldbpagestyle}
\begingroup
\renewcommand\thefootnote{}\footnote{\noindent
This work is licensed under the Creative Commons BY-NC-ND 4.0 International License. Visit \url{https://creativecommons.org/licenses/by-nc-nd/4.0/} to view a copy of this license.
}
\addtocounter{footnote}{-1}\endgroup


\ifdefempty{\vldbavailabilityurl}{}{
\vspace{.3cm}
\begingroup\small\noindent\raggedright\textbf{PVLDB Artifact Availability:}\\
The source code, data, and/or other artifacts have been made available at \url{\vldbavailabilityurl}.
\endgroup
}



\newcommand{\eat}[1]{}
\newcommand{\fullversion}[1]{#1}

\section{Introduction}

\label{introduction}

Numerous software applications utilize SQL to store, manage, and query data within databases. Checking the correctness of application-specific SQL queries by running them on appropriate datasets to check the correctness of the results is an essential component of software testing. 

More recently, there is increasing use of text-to-SQL systems based on LLMs, for data analytics.  Given the possibility of errors (whether due to ``hallucination'' or ambiguity in the textual specification), there is concern about the correctness of the queries.  The creation of effective test datasets is key for detecting errors in generated SQL queries, especially in the common cases where users are not familiar with SQL.

Traditionally, manual test cases are created to test the SQL queries. Query's results on test cases are verified with the expected result set.  However, manually created test cases often do not ensure that all the relevant corner cases are covered.
Automated generation of data for testing SQL queries is thus an important task.  

Errors in formulating a query are modeled as mutations of the query.
A \textit{mutation} of a query is a syntactic variant of the query, for example, one where $<$ in a condition such as $r.A < 5$ is replaced by $<=$.  
A dataset is said to \textit{kill} a mutation of the query if the query and its mutation give different results on that dataset.  Note that some mutations may be semantically equivalent, although syntactically different.

A human tester would take the query specification (in a natural language such as English), and check if the output of the query on each dataset matches the desired output.   

Given a query, the goal of test data generation is to generate datasets that can (ideally) kill all non-equivalent mutations of the query.  Given that queries may be complex, and the number of mutations is much larger, any system targets a particular class of queries and a particular space of mutations.


Early work on generating test datasets to kill mutants includes Agenda \cite{agendatool} and \cite{Tuya}; the most relevant work in this context is the XData system \cite{xdata:2011,xdata:2013,xdata:2015}, which unlike the earlier work implemented a system that generates multiple datasets targeted at different types of mutations.
While XData initially targeted simple select-from-where queries \cite{xdata:2011}, it was subsequently extended to handle a variety of other query features as well as a wider class of mutations \cite{xdata:2015}.   

However, the earlier work on XData had several limitations, in particular in dealing with subqueries.  While single level subqueries were handled, there were challenges in handling more complex queries with subqueries, which are common in the real world.

In this paper, we present a novel data generation approach that is designed to handle complex queries.
As in \cite{xdata:2015}, our approach is based on creating tuples of constraint variables, and arrays of these tuples, and generating constraints on these variables based on the given query.  We then use an SMT constraint solver (Z3) to generate solutions, from which datasets can be directly created.    Multiple sets of constraints are created to generate multiple datasets, to kill a variety of mutations.

The main novel contributions of this paper are as follows:
\begin{enumerate}
    \item We propose a representation of multiset relations using tuple counts, which allows a tuple count of 0 indicating the tuple does not exist; this approach allows the solver to choose the required relation size, up to some limit, rather than have to retry with different sizes.
    \item We propose an approach for representing results of expressions and subqueries as tables, which allows us to handle complex queries in a modular fashion. We call our approach the \textit{result tables approach}.
    \item We describe how to handle non-correlated subqueries using the result table approach, and describe how to handle correlated subqueries by performing decorrelation.  
    \item We present a performance study which shows the effectiveness of our approach on queries for which the earlier XData approach is not as effective.  We also show that it can outperform the state-of-the-art VeriEQL system in showing non-equivalence of query pairs. 
\end{enumerate}

We note that while there has been a significant amount of recent work on proving query equivalence, including the generation of datasets to demonstrate non-equivalence of queries (see Section~\ref{sec:relwork} for details), the body of work related to (non)equivalence depends on the availability of two queries, one of which is the correct query.  Query equivalence testing is thus not applicable for the common use case where a correct query is not already available.  Connections to query equivalence testing are discussed further in Section~\ref{sec:relwork}.

The rest of the paper is organized as follows.  Section~\ref{sec:background} provides background focusing on the XData system.  Section~\ref{sec:overview} provides an overview of our techniques, including the multiset representation.  Section~\ref{sec:query-result-table} describes the result table approach in detail.  Section~\ref{sec:dataset:gen} describes the dataset generation process, including datasets for non-empty results, and other datasets for killing mutations.
Section~\ref{sec:subqueries} describes how we handle subqueries, in particular correlated subqueries.
Section~\ref{sec:relwork} describes related work, while Section~\ref{sec:perf} presents a performance study.


\section{Background}
\label{sec:background}




The XData system \cite{xdata:2011, xdata:2013, xdata:2015} was designed to generate multiple datasets for a given SQL query. Mutations are killed by generating multiple datasets that are targeted at killing mutations within a target space of mutations, which reflect common errors when writing SQL queries. 

The first dataset is designed to generate a non-empty dataset for the original query, which can kill mutations that would generate an empty result on that dataset. Other datasets target one or more types of mutations. 

The dataset generation algorithm of XData works as follows.
\begin{itemize}
    \item Each table in the given schema is modeled as an array of tuples, where each tuple has a variable corresponding to each attribute. 
    \item Constraints are generated on these variables to enforce integrity constraints such as primary keys, foreign keys, etc.  
    \item Further constraints are added, to ensure a non-empty result for the given query.   These constraints are then given to a constraint solver, to generate concrete values.  While an earlier version of XData used the CVC3 solver, the current version uses the Z3 solver.   The values generated by the solver are used to create a dataset.
    \item Further datasets are created, targeting each type of mutation.   The constraints for each of these datasets are similar to the non-empty case constraints, but with modifications to generate a difference in the result of a targeted node in the query tree, from the result of a targeted mutation of that node.  Care is also taken to ensure that the difference at the targeted node also results in a difference in the result of the SQL query.
    \end{itemize}




Consider the following example query:
\begin{verbatim} 
    SELECT R.A FROM R WHERE R.C > 6 ;
\end{verbatim}
The mutations of the comparison are R.C = 6, R.C < 6, R.C >=6 and R.C <= 6.  To kill these mutations, XData would create 3 datasets, which have a tuple where R.C < 6, R.C = 6 and R.C > 6, respectively.   
These datasets also kill missing selection conditions.

XData \cite{xdata:2011} also kills mutations of join type from one of inner, left outer, right outer, and full outer, to any other join type in this set.  Further, since join orders are not specified in SQL, each query has a number of alternative join trees, and mutation of each join node in each of these trees is included in the join mutation space.  The number of such trees is exponential in the number of relations.  


Given the conditions, $r.A = s.B \wedge s.B = t.C$, we can infer the extra condition $r.A = t.C$; the inferred condition allows a larger space of (cross product free) join trees.  In this case, we say that $r.A, s.B$ and $t.C$ are in the same
\textit{equivalence class}.  XData considers mutations across this wider class of equivalent query trees.

To handle join type mutations for a join of n relations $r_1, \ldots, r_n$, XData \cite{xdata:2011} creates datasets such that for each join condition, a dataset ensures one relation $r_i$ lacks a matching tuple on that condition while ensuring there is a collection of tuples that satisfy all the other join conditions. 
Missing join condition mutations are also killed by these datasets.
These polynomial number of datasets help kill non-equivalent mutations in the exponential space of mutations. 



XData also handles
mutations between any one of {LIKE, ILIKE, NOT LIKE, NOT ILIKE} to any other operator from this set, extra and missing group by attributes, and mutations between any of the aggregate functions MIN, MAX, SUM, COUNT and AVG, as well as the DISTINCT variants of these functions.

Prior to our extensions, XData \cite{xdata:2015} handled single-level nested subqueries and the constraint generation approach was hard to extend to more complex queries.  It also had limitations in handling NULL values, including in aggregate operations. Further, it did not consider multiset semantics.  The approach presented in this paper removes these and other restrictions.

\section{Overview of Our Techniques}
\label{sec:overview}


While we follow the same broad approach as the earlier versions of XData \cite{xdata:2011,xdata:2015} for generating multiple datasets targeting different mutations, our system has a greatly improved architecture for representing relations, and for generating the datasets to kill mutations.  We first describe our representation for (multiset) relations, and how integrity constraints are encoded, and then describe how we generate constraints from the given query to generate different datasets to kill mutations.

\begingroup
\let\clearpage\relax
\subsection{Multiset Representation}
\label{multi-set}

Approaches for data generation based on SMT solvers, such as \cite{2010qex} represent a table as a collection (arrays or lists) of tuples where each attribute is a constraint variable.  A drawback of these approaches is that the exact number of tuples should be predetermined. If there are too few or too many tuples, some constraints may not be satisfiable.  To ensure a solution is found, many solvers iteratively retry with different numbers of tuples for each relation, until a solution is found. However, this can be very inefficient, particularly so if there are many relations. 

In contrast, we use an approach of adding a count attribute with each relation, that allows tuples with zero count to be treated as invalid and ignored.
Specifically, we represent the multiplicity of each tuple in a table using the additional integer type attribute CNT. The CNT of each tuple $t_i$ can take any value that is zero or higher. 

A zero value signifies that the tuple $t_i$ is invalid and our constraints are designed to ignore such tuples, using a helper function isValid(t) which returns true if CNT $> 0$ and false otherwise. 
A CNT value of one of greater signifies the multiplicity of the tuple in the table.

Our approach allows the solver to choose a number of tuples for each relation, that can be any value bounded by the number of tuples in the collection (array/list), 
avoiding the high cost of retrying with different numbers of tuples in each relation.
VeriEQL \cite{verieql2024} uses a ``deleted'' boolean value for a similar purpose to our use of 0 multiplicity.





\subsection{Data Types}
\label{data-type}

We follow the XData \cite{xdata:2011,xdata:2015} approach of preprocessing the database schema to get the attribute types and integrity constraints, and generate corresponding SMT constraints.  

As in XData \cite{xdata:2011}, we preprocess sample datasets provided by the user to extract values for different attributes, so we generate datasets whose 
attribute values are realistic/meaningful to users/testers.  

Numeric attributes are modeled as subtypes of integers or real. To model text attributes, enumerated datatypes are used. 
For eg. if we have \verb|dept_name| as an attribute in \verb|department| table then sample values may be provided as follows:
\begin{verbatim}
(declare-datatype DeptNameType
    (Finance History Physics Music))
\end{verbatim}
Attributes belonging to the same equivalence class (i.e. whose attributes are compared) are assigned the same datatype to make them comparable.
Constraints on acceptable values (corresponding to ``check'' constraints) can be defined for each attribute in a table. 

A tuple type is created for each relation, and an array of tuples of that type is then created as illustrated below,
assuming the types IDType, NameType, and DeptNameType are already declared.

\begin{verbatim}
(declare-datatype student_tuple  
    ((tuple (ID IDType) (Name NameType) 
        (DeptName DeptNameType)  (TotCred Int))))
(declare-fun student () (Array Int student_tuple))
\end{verbatim}

To access the attribute from a particular tuple of a table, we use the select construct in Z3. For eg. selecting the name column from the i\textsuperscript{th} student tuple is translated as 
\verb|(Name| \verb|(select student i))|. 

 
Unlike \cite{xdata:2015} we represent NULL for each data type using a single value from the domain. For numeric datatypes, we assign the negative of a large value from the domain of the attribute as NULL. For string attributes, we enumerate one more value along with the domain of the attribute and designate it as a NULL value. 

A function \verb|ISNULL_| \verb|Attr(Attr)| checks if \verb|Attr| value is NULL i.e. is equal to \verb|NULL_Attr|, returns true in that case, otherwise returns false. 
As in Qex, every comparison includes a check to return true only if the values being compared are not null. \footnote{We do not currently support the truth value ``unknown'', which is required for a full implementation of three-value logic.}
Similarly aggregate function calculations skip null values.
We also support the IS NULL clause.


\subsection{Handling Integrity Constraints}

Since multisets are handled by adding a tuple count, primary key constraints are enforced by adding constraints to ensure that 
(i) for every tuple in the relation, the CNT is either 0 or 1,  and 
(ii) if any two tuples in the relation have the same values on the primary key attributes, at least one of them is invalid, i.e. has CNT = 0.

Foreign key constraints are enforced by creating constraints to ensure 
that for each valid tuple in the referencing relation, if the foreign key attribute values are not null, then there is a valid matching tuple in the referenced relation.\footnote{The implementation of both primary and foreign key constraints was slightly different in XData \cite{xdata:2015}, which allowed duplicate tuples and thus represented primary keys as functional dependencies, and avoided existential constraints by allocating a specific tuple in the referenced relation for each tuple in the referencing relation.}  
Not null, domain, and check constraints are also handled by creating appropriate constraints on each tuple in the relation.

\endgroup

\subsection{Result Table Approach}
\label{sec:outline:queryresult}

We create \textbf{result tables} representing the result of subexpressions of a query as well as results of subqueries.   
Given any operation, such as a join or aggregate operation, we create a new table representing the result of the operation, and add constraints to ensure that the contents of the result table match the result of the operation on the input tables.  For a join operation, for example, these constraints ensure that: 
\begin{enumerate}
\item 
For every combination of valid tuples in the input tables that also satisfy the join conditions, there is a matching tuple in the result relation (which we call ``\textit{forward constraints}''), and 
\item Every valid tuple in the result is generated from valid tuples in the input relations that satisfy the join conditions (which we call ``\textit{backward constraints}'')
\end{enumerate}

By creating result tables bottom up from the inner-most/lower levels of a query, we define the result table for the main query. 
Solving for the generated constraints, with an additional constraint that the final result has at least one valid tuple, gives concrete values for each relation which results in a non-empty query result.

Details of the constraints on result tables are described later in Section~\ref{sec:query-result-table}, where we describe how to create constraints to define the results of various operations such as selections/joins, groupby/ aggregate expressions, duplicate-elimination (distinct), set and other operations.   

A major benefit of this approach is that upper layers of the query can treat the subexpression as an input relation, making the approach very modular, and capable of handling complex expressions.

Correlated subqueries create some complications, which we handle as discussed 
later in Section~\ref{sec:subqueries}.

The result table approach forms one of the key differences from XData \cite{xdata:2015} which does not have this concept. 
VeriEQL \cite{verieql2024} also uses a similar concept, although the implementation is different as discussed in Section~\ref{sec:relwork}, and our approach was developed independently.

\subsection{Datasets to Kill Mutations}
\label{sec:killingmutants}

To kill mutations, as in XData \cite{xdata:2015} we first create a dataset that generates a non-empty result as we sketched in the previous section.  
The function \textit{generateDatasetForQuery(q)} which is outlined in Section~\ref{sec:outline:queryresult}, implements the approach.   
The dataset for creating a non-empty result kills all mutations that result in an empty result on the generated dataset.

However, we also have to generate multiple other datasets to kill mutations.
Our algorithm for generating these datasets begins by identifying a set of mutation types and locations for each of the mutation types, and then generates  corresponding datasets, each of which targets one or more mutations of the identified types. 

A \textit{mutation structure} records a mutation location, and a mutation at the identified location. 
For example, we may identify a specific selection node, and consider a mutation of $A=5$ to $A \leq 5$.
We create a collection of all the locations/mutations that we consider, by traversing the query tree and at each node identifying the set of mutations that should be considered at that node, and adding a corresponding mutation structure to the collection.  Joins are special-cased by flattening them, as is done in \cite{xdata:2015}.   

For each mutation structure in the collection, we then need to generate constraints to create datasets targeted at that specific mutation.
The function \textit{generateDatasetForQuery(q, mutstruct)} which we describe later in Section~\ref{sec:datagen:mutation} performs this task.  This function is a modification of 
the function \textit{generateDatasetForQuery(q)}, 
which generates the same constraints for all operations other than the one that is targeted by \textit{mutstruct}, and for that operation it generates constraints designed to kill the mutation.  For example, when considering a particular selection node where $A=5$ is mutated to $A \leq 5$, we would generate a constraint forcing $A < 5$.


\section{Result Table Constraints}
\label{sec:query-result-table}


In this section, we describe our result table approach, which generates constraints to ensure that the table representing a result of an expression has exactly the tuples generated by applying the expression on the input tables.

We first describe how to handle queries with joins and selections, and later extend the approach to handle other operations such as aggregates and set operations.


\subsection{Join Result Table}
\label{sec:join:result}

We describe the creation of constraints for the result of a multi-table join. Hereafter we refer to the join result table as JRT and the tables involved in the join operation as input tables. 

Attributes of JRT include all the attributes from all the input tables along with a CNT attribute. We give unique names to the attributes of JRT: the attribute of $JRT$ corresponding to attribute $A$ from input table $IT$ is given the name $IT\_A$.
The number of tuples in the array for JRT is determined based on the number of tuples in each input relation, along with primary and foreign key constraints, and selection and join conditions, as discussed later in Section~\ref{sec:resulttable:size}. The product of the number of tuples in each input relation is an upper bound, but since most joins are foreign-key to primary-key joins, the join result size can often be determined to be no more than the size of one of the input relations.

For example given relations $r(A,B, CNT)$ and $s(B,C, CNT)$,
consider the result of $r$ inner join $s$ on the join condition $r.B=s.B$.  The result table JRT would have attributes:
\\ \hspace*{5mm} $(r\_A, r\_B, r\_CNT, s\_B, s\_C,  s\_CNT, CNT)$\\
If $r$ has tuples $\{(1, 2, 2), (2, 3, 1)\}$ and $s$ has $\{(2, 4, 3), (2, 7, 2), (2, 9, 0)\}$, then JRT would have tuples
$\{(1, 2, 2, 2, 4, 3, 6), (1, 2, 2, 2, 7, 2, 4)\}$.
The CNT of the tuple is the product of the counts of the input tuples.  Note that the last tuple of $s$ has CNT of 0, indicating it is invalid, so even though its B value of 2 matches that of tuples in $r$, the tuple is not present in the output.

The goal of constraint generation is to ensure that JRT has exactly the required tuples, based on the input tuples.
We consider different forms of the join.

\paragraph{\textbf{Inner Join:}} 
\label{normal-join}

We first create forward mapping constraints, which ensure that every tuple in the result of applying the join operation (along with any selection conditions) on the input tables is present in JRT.  
Listing~\ref{forward-pass-join} shows the generated constraints. For ease of readability, we use the notation $T[i].A$ to refer to the attribute $A$ of the $i$th tuple of the array of constraint tuples representing table $T$.  
The \textit{assert} statements create the constraints that are subsequently solved by the SMT solver.  
In the implementation, we use the SMTLIB2 syntax for representing constraints, but for readability, we use a simpler syntax in this paper.

\begin{algorithm}
\caption{Join Result Mapping}
\label{forward-pass-join}
\begin{algorithmic}[1]
\Function{AttributeMap } {$IT$, $i$, $JRT$, $k$}
    \State Let attributes of $IT$ be $A_1, \ldots\ A_m$
    \State \textbf{return} $(JRT[k].IT\_A_1$ = $IT[i].A_1)$  $\wedge$ $\ldots$\   
    \State \hspace*{1cm} $\wedge (JRT[k].IT\_A_m$ = $IT[i].A_m)$ 
    \State \hspace*{5mm} $\wedge (JRT[k].IT\_CNT = IT[i].CNT)$ 
\EndFunction

\vspace{2mm}
\hrule
\vspace{2mm}

\Function{ForwardMappingForJoin}{$IT_1, \ldots, IT_n, JC$}
\State{// $IT_1, \ldots, IT_n$ are the input tables, $JC$ are the join conditions}
\For{each combination of tuples ($IT_1[i_1],  \ldots, IT_n[i_n]$) \\ 
     \hspace*{2cm} in  ($IT_1 \times \ldots \times IT_n$)}
     \State \textbf{assert}
     \State \hspace{5mm} \textbf{if} $(IT_1[i_1],  \ldots, IT_n[i_n])$ satisfies Join conditions JC  \\
     \hspace*{2cm} $\wedge$ (isValid($IT_1[i_1] \wedge \ldots\ \wedge$ isValid($IT_n[i_n]$)))
    
        \State \hspace{5mm} \textbf{there exists} tuple $JRT[k]$ in JRT 
        \textbf{such that} 
        \State \hspace{15mm} isValid($JRT[k]$)
            \State \hspace{15mm} $\wedge$ \Call{AttributeMap}{$IT_1, i_1$, $JRT$, $k$}
            \State \hspace{15mm} $\ldots$
            \State \hspace{15mm} $\wedge$ \Call{AttributeMap}{$IT_n$, $i_n$, $JRT, k$}

\EndFor
\EndFunction

\vspace{2mm}
\hrule
\vspace{2mm}

\Function{TupleCountMap}{$JRT, k, IT_1, \ldots, IT_n$}
\State \textbf{return}  IsValid($JRT[k]$) $=>$  
\State \hspace*{2mm} $(JRT[k]$.CNT = $JRT[k]$.$IT_1$\_CNT * \ldots * JRT[k].$IT_n$\_CNT)

\EndFunction

\vspace{2mm}
\hrule
\vspace{2mm}

\Function{BackwardMappingForJoin}{$JRT, IT_1, \ldots, IT_n, JC$}
\For{each tuple $JRT[k]$ in JRT }
\State \textbf{assert}

\State \hspace{5mm}\textbf{if} isValid($JRT[k]$) \textbf{then} 

\State \hspace{10mm} Join conditions JC, with attributes $IT_i.A$ renamed to $IT_i\_A$, are satisfied on $JRT[k]$
\State \hspace{10mm} \textbf{and} \Call{TupleCountMap}{$JRT, k, IT_1, \ldots, IT_n$}
\State \hspace{10mm} \textbf{and for} each input table $IT_j$
    \State \hspace{15mm}  \textbf{there exists} tuple $IT_j[j_i]$ 
    \textbf{such that}
        \State \hspace{20mm} isValid($IT_j[j_i]$)
        \State \hspace{20mm} $\wedge$ \Call{AttributeMap}{ $IT_j$, $j_i$,  $JRT, k$}

\EndFor
\EndFunction
\end{algorithmic}
\end{algorithm}
In the backward mapping, we ensure that each valid tuple in JRT has corresponding valid tuples in the input tables that meet the specified join and selection conditions. 


Mapping functions, referred to as AttributeMap, are designed to link each attribute of the input table to its corresponding attribute in the subquery table. These functions are employed in both forward and backward mapping processes.

The backward mapping includes a call to TupleCountMap, which ensures the tuple count of each tuple in the
join output is set to the product of the tuple counts of the input tuples that it was generated from.

Furthermore, to prevent duplicates from being generated among the tuples in JRT, we introduce constraints such that for each pair of tuples in JRT, if the attribute values are identical, at least one of the tuples is invalid i.e. its CNT = 0.




\paragraph{\textbf{Outer Join:}}

Since outer joins are binary operations the constraints are correspondingly on two relations, 
unlike inner joins where our constraints are described for the more general n-ary case. 

Listing~\ref{map-outerjoin} shows the mapping constraints for left outerjoins.  
We modify the forward mapping constraints of the inner join case to handle $IT_1 $ LEFT OUTER JOIN $ IT_2$ ON (\textit{cond}),  as follows.  If a valid tuple $IT_1[i_1]$ from input table $IT_1$ and a valid tuple $IT_2[i_2]$  from input table $IT_2$ together satisfy the condition \textit{cond}, the tuples are mapped to a tuple $JRT[k]$ in the JRT, similar to the inner join case. However, if a valid tuple $IT_1[i_1]$ in $IT_1$ has no match with any valid tuple in $IT_2$ on \textit{cond}, then it is mapped to the outerjoin result tuple similar to the inner join case, but with all attributes of $JRT[k]$ in JRT originating from $IT_2$ set to null using the function AttributeMapNULL.

In backward mapping, the generated constraint ensures that for each valid tuple $JRT[k]$ whose attributes mapped from the input tables satisfies the join conditions, there is a combination of valid tuples $IT_1[i_1]$ and $IT_2[i_2]$ that map to $JRT[k]$. 
Conversely, if the join condition is not satisfied, there is a valid tuple $IT_1[i_1]$ from $IT_1$ that maps to $JRT[k]$, and all attributes in $JRT[k]$ originating from $IT_2$ are NULL and there is no valid tuple in $IT_2$ that satisfies the join conditions with $IT_1[i_1]$.

For the example we saw earlier, compared to the inner join of $r$ and $s$, the left outer join of $r$ and $s$ would have the extra tuple $\{(2, 3, 1, null, null, 1, 1)\}$, since the $r$ tuple $(2, 3, 1)$ does not have a matching $s$ tuple.  Note that the $S\_CNT$ value is 1, rather than 0, to ensure that  $ CNT = R\_CNT * S\_CNT$.

This approach can be adapted to generate a join result table for a right outer join by swapping the roles of input tables $IT_1$ and $IT_2$. The constraints for full outer join can be created by a combination of the left and right outer join cases.  We omit details for brevity.

\begin{algorithm}
\caption{Outer Join Result Mapping}
\label{map-outerjoin}
\begin{algorithmic}[1]
\Function{AttributeMapNULL} {$IT_i$, JRT, $i$}
    \For{each attribute $IT_i\_A$ in JRT coming from $IT_i$ other than TUPLE\_CNT}
        \State  $IT[i].A$ = NULL
    \EndFor
    \State $IT[i].TUPLE\_CNT = 1$
\EndFunction

\vspace{2mm}
\hrule
\vspace{2mm}



\Function{ForwardMappingForOuterJoin}{$IT_1$, $IT_2$, JC}
\State \Call{ForwardMappingForJoin}{$IT_1$, $IT_2$, JC}
\For{for each tuple $t_1$ in input table $IT_1$}
    \If {\textbf{there does not exist }any valid tuple in $T_2$ with join condition true} 
    \State \hspace{5mm} \textbf{there exists} a tuple JRT[k] \textbf{such that}
    \State \hspace{10mm} isValid($k$)
    \State \hspace{10mm} \Call{AttributeMap}{$IT_1$, $t_1$, $k$}
    \State \hspace{10mm} \Call{AttributeMapNULL}{$IT_2$, JRT, $k$}
\EndIf
\EndFor

\EndFunction

\vspace{2mm}
\hrule
\vspace{2mm}

\Function{BackwardMappingForOuterJoin}{JRT, $IT_1$, $IT_2$, JC}
\For{each tuple JRT[k] }
\State \textbf{assert}
\State \hspace{4mm} \textbf{if} isValid($k$) \textbf{and} Join conditions JC, with attributes $IT_i.A$ renamed to $IT_i\_A$, are not satisfied on $k$
    \State \hspace{10mm} \Call{AttributeMapNULL}{$IT_2$, JRT, $k$}
    \State \hspace{10mm}  \textbf{there exists} tuple $t_i$ in $IT_1$
    \textbf{such that}
        \State \hspace{15mm} isValid($t_i$)
        \State \hspace{15mm} \Call{AttributeMap}{$IT_1$, $t_i$, $k$}
        \State \hspace{15mm} \textbf{there does not exists} tuple $t_j$ in $IT_2$ 
        \State \hspace{20mm} isValid($t_j$)

        \State \hspace{20mm} join condition JC is true for $t_i$ and $t_j$
                    \State \hspace{15mm} \textbf{end not exists}

            \State \hspace{10mm} \textbf{end exists}
    \State \hspace{5mm} \textbf{end if}

    \State \hspace{5mm} \textbf{if} isValid($k$) \textbf{and} Join conditions JC, with attributes $IT_i.A$ renamed to $IT_i\_A$, are satisfied on $k$
    \State \hspace{10mm}  \textbf{there exists} tuple $t_i$ in $IT_1$
    \textbf{such that}
        \State \hspace{15mm} isValid($t_i$)
        \State \hspace{15mm} \Call{AttributeMap}{$IT_1$, $t_i$, $k$}
   
        \State \hspace{10mm} \textbf{end exists}
         \State \hspace{10mm}  \textbf{there exists} tuple $t_j$ in $IT_2$
    \textbf{such that}
        \State \hspace{15mm} isValid($t_j$)
         \State \hspace{15mm} \Call{AttributeMap}{$IT_2$, $t_j$, $k$}
        
            \State \hspace{10mm} \textbf{end exists}


   

            
\EndFor
\EndFunction

\end{algorithmic}
\end{algorithm}

\paragraph{\textbf{Semijoin:}}
To generate constraints for a semijoin, we modify the inner join constraints, noting that there are only two inputs $IT_1$ and $IT_2$ and the JRT table has only attributes from $IT_1$. 
In the ForwardMappingForJoin() function we retain the call AttributeMap($IT_1, $ $i_1, JRT, k$) but drop 
the corresponding call for $IT_2$.
We modify the TupleCountMap() function to set $JRT[k].CNT = JRT[k].IT_1\_CNT$ instead of computing the product of counts.  
We modify the BackwardMappingForJoin() function to assert that if isValid($JRT[k]$), then (i) there is a valid tuple $IT_1[j_1]$ for some $j_1$, satisfying the conditions in AttributeMap($IT_1, j_1, JRT, k$) and further (ii) 
there is a valid tuple $IT_2[j_2]$ for some $j_2$ that satisfies the join condition JC with $IT_1[j_1]$.

\paragraph{\textbf{Anti Semijoin:}}
Anti-semijoin returns the tuples from the left input that do not have any matching tuples in the right input.  The tuple count is the same as that for the left input tuple. We modify the forward and backward mapping functions described above for the case of semijoin accordingly.  Details are presented in Listing~\ref{func:killmutants}.  Forward mapping ensures that tuples in the left input ($IT_1$) that do not have a matching tuple in the right input ($IT_2)$ are present in the output.  Similarly the backward mapping constraints ensure that every output tuple corresponds to a tuple from the left input, and further there does not exist any matching tuple from the right input. 

\begin{algorithm}
\caption{Query Result Mapping for Anti Semi Join}
\label{anti-semi-join}
\begin{algorithmic}[1]

\Function{ForwardMappingForAntiSemiJoin}{JRT, $IT_1$, $IT_2$, JC}
\For{for each tuple $t_1$ in input table $IT_1$}
    \If {\textbf{there does not exist }any valid tuple in $T_2$ with join condition true} 
    \State \hspace{5mm} \textbf{there exists} a tuple $JRT_i$ in JRT \textbf{such that}
    \State \hspace{10mm} isValid($JRT_i$)
    \State \hspace{10mm} \Call{AttributeMap}{$IT_1$, $t_1$, $JRT_i$}
\EndIf
\EndFor

\EndFunction

\vspace{2mm}
\hrule
\vspace{2mm}


\Function{BackwardMappingForAntiSemiJoin}{JRT, $IT_1$, $IT_2$, JC}
\For{each tuple $JRT_i$ in JRT }
\State \textbf{assert}
\State \hspace{4mm} \textbf{if} isValid($JRT_i$) 
    \State \hspace{10mm}  \textbf{there exists} tuple $t_i$ in $IT_1$
    \textbf{such that}
        \State \hspace{15mm} isValid($t_i$)
        \State \hspace{15mm} \Call{AttributeMap}{$IT_1$, $t_i$, $JRT_i$}
        \State \hspace{15mm} \textbf{there does not exists} tuple $t_j$ in $IT_2$ 
        \State \hspace{20mm} isValid($t_j$)

        \State \hspace{20mm} join condition JC is true for $t_i$ and $t_j$
                    \State \hspace{15mm} \textbf{end not exists}

            \State \hspace{10mm} \textbf{end exists}
    \State \hspace{4mm} \textbf{end if}
\EndFor
\EndFunction

\end{algorithmic}
\end{algorithm}


\subsection{Selection Result Table}
\label{sec:result:selection}

Selection conditions in the WHERE clause of a query can be added to the join conditions, so we do not create a separate result table for selection operations that occur along with join conditions.  

The case of a query with a single relation, without a join, can be handled by creating forward constraints which ensure that every valid tuple satisfying the selection maps to a valid tuple in the result table, while the backward constraints ensure that every valid tuple in the result table maps to a valid tuple in the input which satisfies the selection conditions.
We omit details, but note that these are a simplification of the constraints generated for joins, by considering a special case with only 1 relation. 

The case of HAVING clause, which imposes a selection condition on an aggregation result is dealt with similarly, by creating a result table for the aggregation and then imposing the selection constraint on that result, as described shortly in 
Section~\ref{sec:agg-result}.

\subsection{Aggregation Result Table}
\label{sec:agg-result}

The aggregation result table represents the result of group by/agg\-regation operations performed on a single input table. The case of aggregation on the result of the join is handled by first creating the result table for the join and then creating the result table for the aggregation using the join result table as input. We denote the aggregation result table as ART.



The attributes of ART include all the attributes in the group by clause, and the aggregation operations.  For eg. for the aggregation operation $SUM(A)$ GROUP BY $B(R)$, the ART has attributes $B$ and $SUM\_A$. Each tuple in ART correspond to a group.

We first add constraints to ensure that no two tuples in ART agree on all group by attributes. 
and then add further constraints shown in Listing~\ref{fm-ART}.
The function ForwardMappingForAgg ensures that the group by value of any valid tuple in the input table is also present in a valid tuple in the ART table. 
The function BackwardMappingForAgg adds constraints to ensure that every valid tuple in ART maps to at least one valid tuple in the input table with the same group by attributes.

\begin{algorithm}\caption{Aggregation Result Mapping}
\label{fm-ART}
\begin{algorithmic}[1]
\ScFunction{ForwardMappingForAgg}{ART, IT, GBAttr}

\State \textbf{for each tuple} $IT[i]$ in input table IT 
\State \textbf{assert}
\State \hspace{2mm} \textbf{if} isValid($IT[i]$)
    \State  \hspace{5mm} \textbf{there exists} tuple $ART[k]$ in ART
    \textbf{such that}
        \State \hspace{10mm} isValid($ART[k]$)
        \State \hspace{10mm} $ART[k].GBAttr = IT[i].GBAttr$
\EndFunction

\vspace{2mm}
\hrule
\vspace{2mm}

\Function{BackwardMappingForAgg}{ART, IT, GBAttr}
\State \textbf{for each tuple} $ART[k]$ in ART 
    \State \textbf{assert}
    \State \hspace{2mm} \textbf{if} isValid($ART[k]$)
    \State \hspace{5mm} \textbf{there exists} tuple $IT[i]i$ in input table IT
    \textbf{such that}
    \State \hspace{10mm} isValid($IT[i]$)
        \State \hspace{10mm} $IT[i].GBAttr = ART[k].GBAttr$
\EndFunction
\end{algorithmic}
\end{algorithm}



Finally, we add constraints to ensure the correct value of the aggregate results in each group. Function CountAggConstraints() in Listing \ref{count-ART}.
specifies the constraints for computing the COUNT aggregate $count(R.A)$ GROUP BY $R.B$.
The function can be easily generalized for grouping by multiple attributes, as well as for aggregation without any group by clause.

Note that if all tuples for a particular group have the attribute A value as NULL, COUNT(A) will be 0.
If the semantics of having the aggregate result as NULL instead of 0 is desired the function can be easily modified to set COUNT\_A to NULL instead of 0.



\begin{algorithm}
\caption{Calculate COUNT(R.A) with group by R.B}
\label{count-ART}
\begin{algorithmic}[1]
\Function{CountAggConstraints}{(ART, IT)}
\For{each tuple $ART_[k]$ in ART}
    \For{each tuple $IT[i]$ in input table IT}
    \State  \textbf{let} aggVal[$t_i$] = \textbf{if} (isValid($IT[i]$) 
    \State \hspace*{5mm}$\wedge\  IT[i].B = ART[k].B $ $\wedge$
      not IsNull($IT[i].A$) 
        \State \hspace*{5mm} \textbf{then}  IT[i].CNT 
        \State \hspace*{5mm} \textbf{else } \textbf{else}  0
    \EndFor
\State \textbf{assert} $ART[k].COUNT\_A =$
\State \hspace*{5mm} $ aggVal[t_1] + aggVal[t_2] + \ldots + aggVal[t_n]$
\EndFor
\EndFunction
\end{algorithmic}
\end{algorithm}

\begin{algorithm}
\caption{Calculate MIN(R.A) with group by R.B}
\label{min-agg}
\begin{algorithmic}[1]
\For{each tuple $ART[k]$ in ART}
    \For{each tuple $IT[i]$ in input table IT}
    
    \State  \textbf{if}  (isValid($i$) \textbf{and} IT[$i$].B = ART[$k$].IT\_B \textbf{ and } IT[$i$].A $\neq$ \text{NULL})
        \State  \hspace{5mm} \textbf{let} aggVal[$i$] = IT[$i$].A
        \State \textbf{else } 
        \State \hspace{5mm} \textbf{let} aggVal[$i$] = MAX\_VAL
    \State \textbf{end if}
    \EndFor
\EndFor
\State \textbf{let} minVal = min(aggVal[$t_1$], aggVal[$t_2$], aggVal[$t_3$]), $\ldots$, aggval[$t_n$]) 
\State \textbf{if} minVal = MAX\_VAL 
    \State \hspace{5mm} ART[$k$].MIN\_A = NULL
\State \textbf{else}
    \State \hspace{5mm} ART[$k$].MIN\_A = minVal
 \end{algorithmic}
\end{algorithm}

We can generate constraints for SUM(R.A) by modifying the COUNT calculation by using ``\textbf{if} not isNull($IT[i].A$) \textbf{then}  $IT[i].A * IT[i].CNT$  \textbf{else} 0'' in place of $IT[i].CNT$.
Furthermore, to handle the case where all the input tuples for a group have NULL value for the aggregate attribute, we also add constraints for computing the COUNT($R.A$), and set the SUM to NULL if COUNT($R.A$) is null.

To compute AVG(R.A) we use $SUM(R.A)/COUNT(R.A)$, if both values are not NULL, and set it to null otherwise.

The MIN and MAX aggregates can be handled in a similar fashion to SUM.
To calculate MIN(R.A), if the values of A for all the tuples of the group are NULL, then MIN(R.A) will be NULL, otherwise, the minimum value of A is defined by the constraints shown in Algorithm~\ref{min-agg}.  Constraints for the calculation of the MAX aggregate are generated similarly.

  



\eat{
\begin{algorithm}
\caption{Calculate MIN(R.A) with group by R.B}
\label{min-agg}
\begin{algorithmic}[1]
\For{each tuple $ART_i$ in ART}
    \For{each tuple $t_i$ in input table IT}
    
    \State  \textbf{if}  (isValid($t_i$) \textbf{and} IT[$t_i$].B = ART[$ART_i$].IT\_B \textbf{ and } IT[$t_i$].A $\neq$ \text{NULL})
        \State  \hspace{5mm} \textbf{let} aggVal[$t_i$] = IT[$t_i$].A
        \State \textbf{else } 
        \State \hspace{5mm} \textbf{let} aggVal[$t_i$] = MAX\_VAL
    \State \textbf{end if}
    \EndFor
\EndFor
\State \textbf{let} minVal = min(aggVal[$t_1$], aggVal[$t_2$], aggVal[$t_3$]),$\ldots$, aggval[$t_n$]) 
\State \textbf{if} minVal = MAX\_VAL 
    \State \hspace{5mm} ART[$ART_i$].MIN\_A = NULL
\State \textbf{else}
    \State \hspace{5mm} ART[$ART_i$].MIN\_A = minVal
 \end{algorithmic}
\end{algorithm}
}




SQL also supports aggregation with DISTINCT clause, for example COUNT(DISTINCT $A$), which removes duplicates before performing aggregation.  To implement this, we can first perform a duplicate removing projection on the group by and aggregated values, as outlined next, and then perform the aggregation.  For example, for an aggregate COUNT(DISTINCT A) group by B (r), we first create a result table for projection of $r$ on on B, A, removing duplicates during the projection, and then use that result table as the input for aggregation.


\paragraph{\textbf{Projection Operation}}
\label{par-projection-operation}
There are two variants of the project operator, one which removes duplicates (corresponding to the use of SELECT DISTINCT) and one that does not (corresponding to plain SELECT in SQL).

In either case, we treat the projection operation as an aggregation operation, grouped by the columns that are projected.  For the case of duplicate removal, the CNT of the output tuples would simply be set to 1, while for the case where duplicates are not removed, CNT is set to the result of the COUNT aggregate function.  

Note that if we simply project out the relevant attributes, the result table may have two tuples whose attribute values are the same, which does not match our multiset representation; these tuples must be merged and their counts added, which is 
implemented by the COUNT aggregate.

\paragraph{\textbf{Having Clause:}}
\label{having-describe}

We next consider the case where there is a constraint on the aggregation result, via a HAVING clause, illustrated by the following example:
\begin{verbatim}
Q1: SELECT R.A FROM R GROUP BY R.A having COUNT(R.B) < 10
\end{verbatim}
We first define the result table for the aggregation result, including any aggregate values that are used in the HAVING clause. 
We can create another result table using the aggregation result table as input, and then impose the selection conditions from the HAVING clause on the aggregate result attribute of that table.  As an optimization, we directly impose the condition on the aggregation results without creating another level of result tables.

It is worth noting that creation of the result table results in a much simpler and cleaner way of handling constraints on aggregation results, compared to \cite{xdata:2015}, which
had a rather complicated way of handling such constraints.


\subsection{Set Operation Result Table}
\label{sec:result:set}

Now we describe how we handle set operators using the result table approach. Set operator queries are of a form Q1 SETOP Q2, where SETOP is one of union, intersection or set difference, with variants ALL or DISTINCT, which preserve or remove duplicates.

We first create result tables for Q1 and Q2 independently. Consider $Q1RT$ and $Q2RT$ be the result tables for queries Q1 and Q2 respectively.  Note that both tables must have the same number/types of attributes, and we assume they have also been renamed to have the same attribute names.

The first step in set operation result table computation is similar to a full outer join of the two queries, with the only difference is that the attributes are stored only once; but in addition, the CNT values from Q1 and Q2 are stored in the result as Q1\_CNT and Q2\_CNT.  If the tuple is present in both Q1 and Q2 with non-zero CNTs in both, both Q1\_CNT and Q2\_CNT would be non-zero.  Otherwise at least one of the two counts would be zero.

The final step computes the CNT value for each tuple, and varies depending on the aggregate operation.
For the case of UNION ALL, the CNT of the result is $Q1\_CNT + Q2\_CNT$, while for UNION, CNT is set to $min((Q1\_CNT + Q2\_CNT), 1$.  For INTERSECT ALL, CNT is set to $min(Q1\_CNT, Q2\_CNT)$ while for INTERSECT, CNT is set $min(Q1\_CNT, Q2\_CNT, 1)$.  For EXCEPT ALL, CNT is set to $max(Q1\_CNT-Q2\_CNT, 0)$, while for EXCEPT the CNT is set to 
$min(max(Q1\_CNT-Q2\_CNT, 0), 1)$.

\eat{
\paragraph{\textbf{Union:}}
We now consider constraints for the union operation. Let USQ denote the result of $P \bigcup R$; USQ contains the projected attributes from both the queries P and R as well as P\_CNT, R\_COUNT, and USQ.COUNT. 

\begin{algorithm}[H]
\caption{Union Result Mapping}
\label{union-RT}
\begin{algorithmic}[1]

\Function{ForwardMappingForUnion}{$IT_1$, $IT_2$, USQ}

\For{each tuple $t_i$ in $IT_1$ }
    
     \State \textbf{assert}
     \State \hspace{5mm} \textbf{if} isValid($t_i$) 
    
        \State \hspace{10mm} \textbf{there exists} tuple $usq_i$ in USQ 
        \textbf{such that} 
        \State \hspace{15mm} isValid($usq_i$)
        \State \hspace{15mm} \textbf{and}
    
            \State \hspace{15mm} \Call{AttributeMap}{$IT_1$, $t_i$, $usq_i$}
        
        \State \hspace{10mm} \textbf{end exists}
\EndFor
\For{each tuple $t_j$ in $IT_2$ }
    
     \State \textbf{assert}
     \State \hspace{5mm} \textbf{if} isValid($t_j$) 
    
        \State \hspace{10mm} \textbf{there exists} tuple $usq_i$ in USQ 
        \textbf{such that} 
        \State \hspace{15mm} isValid($usq_i$)
        \State \hspace{15mm} \textbf{and}
    
            \State \hspace{15mm} \Call{AttributeMap}{$IT_2$, $t_j$, $usq_i$}

        \State \hspace{10mm} \textbf{end exists}
\EndFor

\EndFunction

\vspace{2mm}
\hrule
\vspace{2mm}

\Function{BackwardMappingForUnion}{USQ, $IT_1$, $IT_2$}

\For{each tuple $usq_i$ in $USQ$ }
    
     \State \textbf{assert}
     \State \hspace{5mm} \textbf{if} isValid($usq_i$) 
    
        \State \hspace{10mm} \textbf{there exists} tuple $t_i$ in $IT_1$ 
        \textbf{such that} 
        \State \hspace{15mm} isValid($t_i$)
        \State \hspace{15mm} \textbf{and}
    
            \State \hspace{15mm} \Call{AttributeMap}{$IT_1$, $t_i$, $usq_i$}
        
        \State \hspace{10mm} \textbf{end exists}

        \State \hspace{10mm} \textbf{or}
        \State \hspace{10mm} \textbf{there exists} tuple $t_j$ in $IT_2$ 
        \textbf{such that} 
        \State \hspace{15mm} isValid($t_j$)
        \State \hspace{15mm} \textbf{and}
    
            \State \hspace{15mm} \Call{AttributeMap}{$IT_2$, $t_j$, $usq_i$}
        
        \State \hspace{10mm} \textbf{end exists}
\EndFor
\EndFunction
\end{algorithmic}
\end{algorithm}

Since the UNION operator does not contain duplicates, we enforce the constraints that there are no duplicates on attributes other than CNT (as described in Section~\ref{multi-set}) and that CNT = 1. 

To support UNION ALL, which is the multi-set union, instead of constraining CNT = 1, 
we set it to the CNTs of P if the tuple is absent in R, to the CNT of R if it is absent in P, and to the sum of P.CNT and R.CNT if it is present in both P and R.   

For brevity, we omit details of how intersection and set difference are implemented, but they follow a similar approach to that for union.
}


\section{Dataset Generation}
\label{sec:dataset:gen}

We now provide details of our algorithms for generating datasets, starting with generation of non-empty datasets, followed by datasets for killing mutations, and then outline how to decide the result table size.

We assume in this section that subqueries have been decorrelated and provide an overview of how to handle decorrelated subqueries.  More details on handling subqueries, including how to decorrelate subqueries, is discussed subsequently in Section~\ref{sec:subqueries}.

\subsection{Dataset for Non-Empty Results}
\label{sec:datagen:non-empty}






The overall algorithm for generating a non-empty dataset is shown in Listing~\ref{non-empty}.
First, the query tree is preprocessed, and decorrelation is carried out on all correlated subqueries.

Procedure generateDatabaseConstraints() is invoked to generate SMT constraints from database integrity constraints, and the procedure generateConstraintsForQuery(q) is invoked to generate constraints for the query. 
The procedure
generateDatasetUsingConstraints() then solves the constraints using the SMT solver (Z3) to get a non-empty dataset.

\begin{algorithm}
\caption{Data generation algorithm}
\label{non-empty}
\begin{algorithmic}[1]

\State \textbf{procedure} {generateDatasetForQuery(query q)}
\State \hspace{5mm} decorrelateQueryTree(q)
\State \hspace{5mm} generateDatabaseConstraints()
\State \hspace{5mm} generateConstraintsForQuery(q) 
\State \hspace{5mm} generateDatasetUsingConstraints() /* Calls Z3 solver
\State \hspace{10mm}  on constraints generated */
\State \textbf{end procedure}


\State \textbf{procedure} {generateConstraintsForQuery(query q)}

\State \hspace{5mm} \textbf{for all} subqueries q' in q
\State \hspace{10mm} generateConstraintsForQuery(q')
\State \hspace{5mm} 
replaceSubqueriesByResultTables() 
\State \hspace{5mm} 
createJoinResultConstraints(q) /* Creates JRT /* 
\State \hspace{5mm}  createAggregateResultConstraints(q) /* Creates ART */ 
\State \hspace{5mm} 
createHavingClauseConstraints(q) 
\State \hspace{5mm} 
createSetConstraints(q) 
\State \hspace{5mm} 
createNonEmptyConstraints(q) 

\State \textbf{end procedure}



        

\end{algorithmic}
\end{algorithm}

Procedure generateConstraintsForQuery(q) first calls itself recursively on the (already-decorrelated) subqueries if any are present, and then generates constraints to define the result table of the top level of $q$.  The recursive calls to the subqueries create the constraints defining the result tables for each of the subqueries. 
Each subquery is then replaced by its result table in the main query, with additional constraints added to join the subquery result table with the main table; details of how to do so are described later in Section~\ref{sec:subqueries}. 
 
To define the constraints for the result table of SQL query $q$, the from and where clauses of $q$  are handled by invoking createJoinResultTable(), which adds constraints to create a join result table (JRT) based on the join type as discussed in Section \ref{sec:join:result}.
Subsequently, if aggregation is present, createAggregateResultTable() creates an aggregate result table (ART) on the join result table.
If the query has a SELECT DISTINCT clause, it is handled in the same way as a group-by/aggregation operation, with all attributes in the select distinct clause treated as group-by attributes, and with no aggregation.
If a HAVING clause is present, a result table is added for it, as discussed earlier. 
Next, if there is any set operation at the top level of the query, it is handled by generating appropriate constraints.
Finally, constraints are added to ensure a non-empty query result by requiring that there exists at least one valid tuple in the query result table.

\subsection{Datasets for Killing Mutations}
\label{sec:datagen:mutation}

Listing~\ref{func:killmutants} shows how datasets are generated for killing mutations.  First, the function \textit{collectMutationStructs(q)} is called to create a set of mutation structs for the query.  This procedure traverses the query tree, and at each node it decides on what mutations are to be considered, and adds the (node, mutation) pair to the collection of mutation structs.

The procedure generateDatasetForQuery(q, ms) is a variant of generateDatasetForQuery(q), which generates a dataset targeted at killing the specified mutation at the specified node. For many of the mutations, this is done by adding constraints to ensure the unmutated node gives an empty result, while the mutated node gives a non-empty result.  

\begin{algorithm}
\caption{Dataset generation to kill mutants}
\label{func:killmutants}
\begin{algorithmic}[1]

\State mutationStructure /* Contains type, mutationLocation, mutatedNode, queryBlock, isExpired */
\State \textbf{procedure} generateDatasetToKillMutations(q)
\State $msSet$ =  collectMutationStructs($q$)
\State \textbf{for each}  mutation struct $ms$ \textbf{in} msSet
    \State \hspace{10mm} generateDatasetForQuery($q, ms$ ) 
\State \textbf{end procedure}

\end{algorithmic}
\end{algorithm}










We generate datasets to kill mutations of types such as selection mutations, join condition mutations, missing join conditions, string condition mutations, extra group by and partial group by mutations, and aggregate function mutations by creating appropriate constraints, following the approach described in \cite{xdata:2011,xdata:2015}.

We exclude the details for brevity, but note that
the technique for killing join vs.\ outerjoin mutations creates a small number of datasets that are tailored to kill an exponentially larger number of mutations across alternative join trees for the same query.
While we use the same technique, our implementation has been revamped to clean it up, and to make it work seamlessly with the result table approach.
In the rest of this section we discuss methods to kill additional mutation types beyond those considered in \cite{xdata:2011,xdata:2015}.

\paragraph{\textbf{Distinct Clause Mutations:}}
To catch mutations that drop or add a DISTINCT clause in the SELECT clause, we create a dataset where at least one result table tuple has $ CNT > 1$.



\paragraph{\textbf{Having Clause Condition Mutations:}}
Since the Having clause is represented as a selection on top of aggregation, we use the existing XData techniques for catching selection condition mutations.


\paragraph{\textbf{Non-Equi Join Condition Mutations:}}
For a join condition A relop B, where A and B are attributes and relop is one of the comparison operators ($<, <=, >, >=$ we generate three datasets by mutating the condition to $<$, $>$ and $=$, similar to selection condition mutations.
These three datasets kill all the relop mutations and missing join mutations. These datasets also differentiate between inner and outer joins on the given condition. 

\paragraph{\textbf{Where Clause Connective Mutations:}}
For EXISTS/NOT EXISTS subqueries, the mutation between the two connectives is considered and datasets generated for both cases, one with EXISTS connective and the other with NOT EXISTS connective. 
Using these datasets, mutants with missing or extra subquery are also killed as such mutants will produce empty results on either the EXISTS or NOT EXISTS dataset. Similar datasets are generated for IN and NOT IN.

\paragraph{\textbf{Scalar Subquery Connective Condition Mutations:}}
On performing decorrelation as discussed in  Section~\ref{sec:subqueries}, correlation conditions in the scalar subquery and outer query become conditions for SEMI JOIN between the result table for the subquery and the input tables in the outer query. The join condition mutations are discussed in Section~\ref{sec:background}. 

We note that mutations of the ORDER BY clause cannot be caught using datasets, but equivalence of different order by clause attribute sequences can be checked
as described in \cite{xdata:2019}.

\subsection{Deciding Maximum Number of Tuples}
\label{sec:resulttable:size}

As described earlier, by allowing the tuple count CNT of a tuple to be 0, and treating such tuples as invalid in our constraints, the constraint solver is enabled to generate solutions with the number of tuples in a relation ranging from 0 to the maxiumum size of the array representing each relation.  
However, that still leaves the question of what the size of the array, i.e.\ the maximum number of tuples, should be for each relation.

For each base relation, we decide on the number of tuples as follows.  For each occurrence of the relation in the query, we add one tuple.  Further, if a relation $r$ with $n$ tuples has a foreign key referencing relation $s$, we ensure $s$ has at least $n$ tuples,
i.e. we use the maximum of the sizes of all referencing relations.  

If there are any constraints on an aggregate on the relation, we follow the \cite{xdata:2015} approach to 
determining the number of tuples required, based on the constraints; for example if there is a constraint $sum(r.A) >= 500$, along with a constraint $r.A <= 100$, we will require at least $5$ tuples to be present.
Finally, we note that an aggregation constraint on a result table may force the result table to have at least some number of tuples.  We push this requirement down to the input relations from which the result table is generated.

\begin{algorithm}
\caption{Steps to determine maximum number of tuples}
\label{maxtuples:join}
\begin{algorithmic}[1]
\Function {MaximumJoinSize}{JRT, $IT_1$, $\ldots, IT_n$}

\State \textbf{for each relation} $IT_i$ 
\State \hspace*{5mm} M($IT_i$) = $s_i$ (i.e. number of tuples in $IT_i$)

\State \textbf{for} all selection conditions of the form (primary key of $IT_i$ = constant)
\State \hspace{5mm} M($IT_i$) = 1

\For{all join conditions}  
\State  \textbf{if} join condition is (primaryKey of $IT_i$ = primaryKey of $IT_j$)
    \State \hspace{5mm} $M$(smaller input table) = size (smaller input table)
    \State \hspace{5mm} $M$(other table) = 1
\State \textbf{else if} join condition is equality on primaryKey of $IT_j$)
\State \hspace{5mm} M($IT_j$) = 1 /* join cannot increase num.\ tuples */
\State maxNoOfTuplesInJRT = $M(IT_1)\times M(IT_2$) 
\State \textbf{else if} join condition is (non primarykey of $IT_i$ 
\State \hspace*{5mm} = non primarykey of $IT_j$)
\State \hspace{5mm} $M(IT_i) = 2 $ where $IT_i$ is the smaller relation. 
\State \hspace*{5mm} /* Above is a heuristic */
\EndFor
\State \textbf{return} ($M(IT_1) * \ldots * M(IT_n)$)
\EndFunction

\end{algorithmic}
\end{algorithm}


Next, we consider how to determine the size of result tables based on the sizes of the input relations.

\noindent\textbf{Join Result Table:}
Consider $IT_1$ and $IT_2$ are the input tables for join result table JRT, with maximum number of tuples as $m$ and $ n$ respectively.  Then the maximum number of tuples in JRT can in the worst case be $m*n$.  However, in practise the number of tuples is much lower, due to selections as well as because joins are almost never cross products, and are most commonly foreign-key to primary-key joins.   We can get better size bounds as described below.


\fullversion{
The function maxJoinSize shown in Listing~\ref{maxtuples:join} in the Appendix calculates the maximum number of tuples in a join result table.
}

To compute the maximum number of tuples in the join of $n$ relations, we first compute a value $M(IT_i)$ for each relation $IT_i$ as described below, and then multiply these numbers to get the size bound.  The value $M(IT_i)$ is initialized to the number of tuples in $IT_i$, for all $i$.  If there is a selection condition of a form \{primary/unique key of $IT_i$ = constant \}, we set $M(IT_i) = 1$. 
If the join condition equates the primary/unique key of input table $IT_i$ to non-primary/unique key attrbutes of another relation, we set $M(IT_i) = 1$, since the join condition
can map each tuple of the other table to at most 1 tuple of $IT_i$
If a join condition equates the primary/unique keys of two input tables, we set $M(IT_i)=1$ for the larger of the two tables, since in this case the result cannot be larger than the smaller of the relations.

If the join condition involving two relations does not involve primary/unique keys,
we heuristically reduce the size of JRT as follows: if $M(IT_i) > 2$ for both relations, we set $M(IT_i)=2$ for the smaller $M(IT_i)$.
As a further heuristic, we cap the size of the result table to some maximum value, to avoid problems with solver timeouts with large relation sizes.

We use the same bound for outerjoins also since the size bounds assume every tuple has a matching tuple for each join condition.
For semijoins, the result table size bound is the number of tuples in the left hand side input.


\noindent{\textbf{Aggregate Result Table}}:
The maximum number of tuples in the aggregate result table can be equal to the number of tuples in the input table if each tuple in the input table has a different group by attribute values. Lesser tuples are possible in ASQ, based on the number of groups present. 

\noindent{\textbf{Set Operator Result Table}}:
For set operator result tables, we set the maximum number of tuples in the result table equal to the sum of the number of tuples in the input tables for union, and to the minimum for intersect, and to the left input table for except.

\subsection{Handling String Types}
\label{string-type}
Although Z3 supports solving of string constraints using either of two supported solvers, we found that the performance of these solvers to be very slow for handling string constraints such as comparison and like clause matching.  Therefore we continue to use the XData string implementaion from \cite{xdata:2015}, where the string constraints on an attribute are used to create a set of string values which are then used as the domain for that attribute.  

String matching constraints using the LIKE clause (where the pattern is assumed as in \cite{xdata:2015} to be a constant value), as well as comparisons between string attributes and constraints are handled by using the string solver from \cite{xdata:2015} to generate values that can satisfy the constraints.  In \cite{xdata:2015} both comparison and LIKE constraints (e.g. $r.A < $ 'Bio' and $r.B$ LIKE "C\_") are both replaced by equality constraints with satisfying values generated by the string solver (e.g. $r.A = $ "A"   and $r.B =$ "Ca").
Our implementation handles LIKE constraints in the same way, but handles comparisons by translating them into integer comparisons as described below, allowing the solver flexibility in selecting the assigned values.

Since Z3 does not internally support string comparison, we define a function to map the pre-computed string values to distinct integers, satisfying the lexicographic sort order, and replace a comparison of the form $x < y$ by 
$attrmap(x) < attrmap(y)$ where attrmap is the above-mentioned mapping function for the attribute domain.

Our implementation maps string constrants to values of a datatype for that attribute, as seen in our earlier example defining the type \verb|DEPT_NAME|.  This approach allows the domain of the attribute to be constrained to predefined set of strings, without requiring  the constraint to be declared for each variable of that attribute.
Thus, mapping of strings to integers, is replaced by a function that maps values of the attribute datatype to integers.

\begingroup
\let\clearpage\relax
\section{Handling Nested Subqueries}
\label{sec:subqueries}

The result table approach is used to handle multi-level nested subqueries by treating each subquery as an independent query. Thus, the result of each subquery is represented by its result table, which is then used as an input to the higher-level query that contains the nested subquery.  However, there are several details to be handled, in particular issues in handling correlated subqueries.

\eat{
\begin{figure}
  \centering
  \includegraphics[width=\linewidth]{figures/querytable.png}
  \caption{An illustration of a query table formation on query with 2 levels of nested subqueries} 
  \label{fig:querytable}
\end{figure}
}

In this section we describe how to handle nested subqueries in the FROM clause, and in the WHERE clause with various types of connectives.  
Our approach to decorrelate nested subqueries is based on decorrelation techniques described in \cite{Neumann}.

\subsection{From Clause Subqueries:}
\label{from-clause}

From clause subqueries without correlation are handled in a straightforward manner by defining a result table for the query, and replacing the subquery by the result table when generating constraints for the outer query.
From clause subqueries cannot include correlation variables from the current query, except by the use of the LATERAL clause.  

Although not yet implemented, 
lateral-clause correlation variables can be handled similar to correlation variables in WHERE clause subqueries, except instead of a semijoin we would perform a join of the result table for that subquery.

\subsection{Where Clause Subqueries}
\label{sec:subquery:where}

The mechanism for handling where clause subqueries vary based on the subquery connective, such as EXISTS, IN, etc., and on the use of correlation variables.

\eat{
\paragraph{\textbf{Exists Connective Without Correlation:}}
\label{exist-wo-corr}
Consider a query with a single-level nested where clause subquery with an EXISTS connective, and no correlation variables. 
\begin{verbatim}
Q3:
    SELECT R.A, R.B FROM R                --> Level 0
    WHERE EXISTS (SELECT S.A  FROM S)     --> Level 1
\end{verbatim}
We transform the subquery containing the EXISTS connective into a FROM clause subquery, which then employs a semijoin with the other input tables of the outer query, as described below:\footnote{SQL implementations include semijoin as an internal operation, although they do not typically support the semijoin syntax in SQL.  We use the syntax only for notational convenience; our approach does not depend on the database systems support for semijoin since we only use it as a temporary step towards creating constraints for data generation, and do not actually execute semijoin on the database.}
\begin{verbatim}
Q3':
    SELECT R.A, R.B                       --> Level 0
    FROM R SEMI JOIN (SELECT S.A FROM S)  --> Level 1 
\end{verbatim}
We can handle the above query by creating a result table for the subquery added to the from clause and replacing the subquery by the result table, as described in \ref{from-clause}.



\paragraph{\textbf{Exists Connective:}
\label{exist-with-corr}}
}

\paragraph{\textbf{Exists Connective:}
\label{exist-with-corr}}

Our approach to handling correlation variables is based on the existing approaches for decorrelation of nested subqueries; see for example \cite{Neumann,db-book,sqlserver-decorrelation}.
We illustrate the approach via an example.



\begin{verbatim}
Q1: SELECT R.A, R.B  FROM R               --> Level 0
    WHERE EXISTS (                 
        SELECT S.A FROM S                 --> Level 1
        WHERE S.B = R.A)
\end{verbatim}

The above query can be decorrelated by transforming it to the following query.

\begin{verbatim}
Q1': SELECT R.A, R.B                      --> Level 0
     FROM R SEMI JOIN              
        (SELECT S.A, S.B FROM S) AS SQ1   --> Level 1  
        ON SQ1.B = R.A;
\end{verbatim}

The rewritten query now has the WHERE clause subquery replaced by a FROM clause subquery without correlation, with a semijoin operation.
Note that the attribute used in the correlation condition has been added to the SELECT clause list for the inner query, and the correlation condition S.B=R.A has become the semijoin condition.

Constraints for the transformed query are generated as described earlier for FROM clause subqueries.


Next consider the following query with aggregate in the outer query with an attribute correlated with the inner query. 

\begin{verbatim}
Q2:  SELECT R.A FROM R                  --> Level 0 
     WHERE EXISTS (
        SELECT SUM(S.A) FROM S          --> Level 1
        WHERE S.B = R.B)
\end{verbatim}

We can decorrelate this query as follows.
\begin{verbatim}
Q2':  SELECT R.A FROM R                 --> Level 0
      SEMI JOIN (                       --> Level 1
        SELECT  S.B, SUM(S.A)  FROM S
        GROUP BY S.B) as ASQ1
      ON (ASQ1.B = R.B) 
\end{verbatim}
Note that to make the column S.B available to the semijoin, given that the subquery has an aggregation operation, we had to add the column to a newly added GROUP BY clause in addition to the SELECT clause (if there was already a GROUP BY clause, we would add S.B to the existing columns in the GROUP BY clause).\footnote{ In this toy example the aggregated value in the subquery is not used, and we can actually further optimize the query to remove aggregation, but there are other cases of subqueries where the aggregated values are used, and the decorrelation approach will be applicable even in such cases.}

Each group above corresponds to an invocation of the subquery with a specific value of R.B from the outer query.  However, the approach of adding S.B to the GROUP BY clause does not work if the comparison is
not a simple equality.  For example, consider the 
following query.

\begin{verbatim}
Q3:  SELECT R.A FROM R                   --> Level 0 
     WHERE EXISTS (
        SELECT SUM(S.A) FROM S           --> Level 1
        WHERE S.B < R.B)
\end{verbatim}

To decorrelate this query, we have to simulate the invocation with different values of R.B by adding 
a join with a new subquery that generates  the distinct values of the correlation variable R.B that the original subquery would have been invoked with.  In general, a superset of the invoked values is also acceptable since it would generate the same final result.


\begin{verbatim}
Q3':  SELECT R.A FROM R                  --> Level 0
      SEMI JOIN (                        --> Level 1
        SELECT  R1.B , SUM(S.A) 
        FROM S, (SELECT DISTINCT R.B FROM R) AS R1
        WHERE S.B < R1.B 
        GROUP BY R1.B) as ASQ1
      ON (ASQ1.B = R.B) 
\end{verbatim}
To make these values of R.B available in the result of the decorrelated subquery, we have added it to the GROUP BY clause and the SELECT clause of the decorrelated subquery, and further add a semijoin on R.B = ASQ1.B.

\paragraph{\textbf{Not Exists Connective:}}

Consider the following query.

\begin{verbatim}
Q4: SELECT R.A, R.B   FROM R              --> Level 0
    WHERE NOT EXISTS (                 
        SELECT S.A FROM S                 --> Level 1
        WHERE S.B = R.A)
\end{verbatim}

We decorrelate this subquery with the help of anti-semijoin as follows:

\begin{verbatim}
Q4': SELECT R.A, R.B                       --> Level 0
     FROM R ANTI SEMI JOIN              
        (SELECT S.A, S.B FROM S) AS SQ1    --> Level 1  
        ON (SQ1.B = R.A);
\end{verbatim}

Note that S.B has been added to the SELECT clause of the
subquery to make it available for the anti-semijoin.

\paragraph{\textbf{Multi-Level Nesting of Subqueries:}}
\label{exist-multilevel}

Now let us consider the case of multi-level nested WHERE clause subqueries where the correlation variables used in a subquery come from some level higher than the immediately higher level query.



\begin{verbatim}
Q5: SELECT R.A, R.B FROM R                --> Level 0
    WHERE EXISTS (                 
        SELECT S.C FROM S                 --> Level 1
        WHERE S.C = R.C AND EXISTS (
            SELECT * FROM T               --> Level 2
            WHERE T.A = R.A AND T.B = S.B))
\end{verbatim}

In the example above, the Level 2 subquery has a correlation variable R.A which is from Level 0.  
The correlation variable R.A is not available for implementing the semijoin condition when we decorrelate the Level 2 subquery into Level 1.

The query can be decorrelated as follows:
\begin{verbatim}
Q5': SELECT R.A, R.B                       --> Level 0
     FROM R 
     SEMI JOIN                           
        (SELECT SQ3.A, SQ3.C
         FROM (SELECT S.B, S.C, R.A        --> Level 1
               FROM S, (SELECT DISTINCT R.A FROM R)
              ) as SQ3
            SEMI JOIN                   
              (SELECT *  FROM T) as SQ2    --> Level 2 
            ON (SQ2.B = SQ3.B AND SQ2.A = SQ3.A)
         ) as SQ1 
     ON (SQ1.C = R.C AND SQ1.A = R.A)
\end{verbatim}

Note that the correlation variable R.A has been added to
the enclosing query at Level 1 by adding a join (crossproduct) with R at Level 1.  We can then use R.A when decorrelating the Level 2 subquery, as well as in a semijoin condition when decorrelating the Level 1 subquery.

This models the situation where the correlation value is passed from the outer to the middle level, and then from there to the inner level subquery. 
A more formal description along with further optimizations may be found in \cite{Neumann}.






\paragraph{\textbf{IN/NOT Connectives}}

Consider the following IN/NOT IN queries
\begin{verbatim}
Q6: SELECT * FROM R WHERE R.B IN (SELECT S.B FROM S);
Q7: SELECT * FROM R WHERE R.A NOT IN (SELECT S.B FROM S);
\end{verbatim}
The decorrelated version of the two queries is as follows:
\begin{verbatim}
Q6': SELECT * FROM R SEMI JOIN S ON (R.A = S.B)
Q7': SELECT * FROM R ANTI SEMI JOIN S
      ON (R.A = S.B OR R.A IS NULL OR R.B IS NULL)
\end{verbatim}
The ANTI SEMI JOIN condition in the decorrelated query allows S.A and S.B to be NULL.  This is due to a subtle distinction between NOT IN and NOT EXISTS due to null values.
If R.A or any S.B is null, the NOT IN would fail since we cannot be sure there is no matching tuple, whereas rewriting it without the null check (whether using an ANTI SEMI JOIN as above, or using NOT EXISTS) would result in an empty subquery result even with nulls, and the corresponding R tuple would appear in the result.

Subqueries with ANY connective (e.g. $=$ ANY, or $<$ ANY) can be decorrelated to a semijoin, whereas the $=$ ALL connective can be translated to an anti-semijoin with the negative of the condition (e.g., $=$ is replaced by $!=$).  We omit details due to lack of space.

\paragraph{\textbf{Scalar Subqueries}}
\label{scalar-subquery}

Consider the following query. 
\begin{verbatim}
Q8:  SELECT S.A FROM S 
     WHERE S.B = (SELECT R.B FROM R)
\end{verbatim}

We transform the above scalar subquery into a FROM clause subquery with semijoin as follows:

\begin{verbatim}
Q8':  SELECT S.A 
      FROM S SEMI JOIN (SELECT R.B FROM R) 
          ON (R.B = S.B) ;
\end{verbatim}
The condition linking the scalar subquery is turned
into the semijoin condition, similar to how correlation 
variables are handled.
Scalar subqueries in the SELECT clause can be handled by using a join instead of a semijoin.

The above transformation assumes that the scalar subquery never returns more than 1 answer; if it returns more than 1 answer, there would be a runtime exception. Decorrelated queries cannot throw such runtime exceptions.

\begin{verbatim}
Q9:  SELECT S.A FROM S 
     WHERE S.B = (SELECT COUNT(*) FROM R)
\end{verbatim}
Count bug is scalar subqueries appear for the inner queries with COUNT(*) projected as in Q9.
When there is no tuple in the group matching the conditions COUNT(*) will be 0 generating incorrect results. To avoid this, the left outer join is used instead of semijoin while decorrelating the query.


\endgroup


\section{Query Equivalence Checking}
In section \ref{sec:dataset:gen}, we discussed our approach to generate the datasets to kill mutations. To compare the results of the queries on each dataset, we use a query with symmetric difference set operator to create another query Q = "Q1 symmetric difference Q2". The symmetric difference of two sets, A and B, returns elements that are in exactly one of the sets, but not in both. If the result of Q is empty on a dataset, it means the queries return the same results otherwise the mutation is killed. This approach gives rise to the idea of using symmetric difference operator results on original and mutant queries to check equivalence.

In sections \ref{sec:query-result-table}, we discussed the result of set operators with 2 queries (left query and right query) as inputs. The final step for the set operator result computes the CNT value for each tuple, and varies depending on the aggregate operation.
For the case of symmetric difference, the CNT of the result is $MAX(|Q1\_CNT - Q2\_CNT|, 0)$.

In addition, we assert that at least one tuple exists in the symmetric difference result table. If the model for such constraints is satisfiable then it implies that queries are non-equivalent. If the model is unsatisfiable, that implies either of the following things: 1) The queries are equivalent 2) The number of tuples is not enough. Hence this approach allows us to perform a bounded check for query equivalence. In the future, this approach can be tested and compared against state-of-the-art query equivalence tools.

\section{Related Work}
\label{sec:relwork}

The AGENDA system \cite{agendatool} was one of the early tools for test data generation.  It generates test data for an application, given as input the database schema, and sample value files.  However, the data generated is however query agnostic, and provides no guarantees about killing mutants, and may not even provide non-empty results for a given query.  

Qex 
\cite{2010qex} is a tool for generating input tables to ensure non-empty results for a given SQL query, by generating constraints and solving them using the Z3 SMT solver \cite{z3:2008}.  However, the datasets generated by Qex are not targeted at killing mutations.  

Tuya et al. \cite{Tuya} provide a classification of a space of mutations, and describe techniques to kill some of these mutations but do not provide a system to automatically generate test data to kill these mutations. 



Generated datasets may also be used to test queries written by students, or by a text-to-SQL system; in such a case the testing can be automated by comparing the result of a given correct (``golden'') query with the query being tested.   Query equivalence testing is an alternative to data generation for this use case.

Work on query equivalence testing is closely related, although there are significant differences.
Early work on query equivalence handled restricted classes of queries such a conjunctive queries.  
Cossette \cite{cossetteplus:2018} represents tuples with multiplicity, but does not use an SMT solver and thus has limitations in showing equivalence.
Recent results on query equivalance testing, such as \cite{spes:2022,sqlsolver,verieql2024,verifySQLCVC5},
have significantly extended the class of queries and integrity constraints considered.     
\cite{verifySQLCVC5} uses a second order theorem prover to check query equivalence, and thus handles relations of arbitrary size, but is limited in the class of queries it can handle.
These systems attempt to prove equivalence, but if the queries are not equivalent, some of them such as \cite{verifySQLCVC5,sqlsolver,verieql2024}, attempt to generate datasets that are witness to non-equivalence.



A key limitation of the query equivalence tools is that they cannot be used in case a correct query is not available, and the goal is to generate datasets to check if a given query is correct. 
One could conceivably enumerate mutations of a given query, and use the query equivalence checking tool to generate datasets to prove non-equivalence for each of the mutants. However, the number of such mutations is very high (exponential in the query size as described in \cite{xdata:2011}) which will generate a very large number of datasets.  In contrast, the XData approach to data generation is able to generate a smaller number of datasets that can kill a much larger class of mutations.

The state of the art in query equivalence verification tools, which handles a larger class of queries than earlier tools, is VeriEQL \cite{verieql2024}.   There are some similarities between our result table approach and the query representation approach of \cite{verieql2024}.  In addition to the obvious difference of not supporting generation of test datasets given a single query, there are further differences: 
unlike us, VeriEQL does not handle correlated nested queries, and has other limitations such as not handling commonly used data types such as strings, dates etc. 
We present a performance comparison of our approach with VeriEQL in Section~\ref{sec:perf} which demonstrates the benefits of our approach.



Support for result tables was added earlier to XData as described in \cite{mtp:ravishankar:temptable}, where the tables were referred to as ``temporary tables''.  However, only simple binary joins were handled.     
VeriEQL \cite{verieql2024} also creates relations for the results of operations, and adds constraints to ensure the result corresponds to the inputs.  For the case of joins, the output relation size is the cross product of the input relation sizes, with input tuples mapped to specific locations on the output table; tuples that do not satisfy the join condition are marked as deleted. 
The VeriEQL approach results in large relation sizes, even for common case of foreign-key joins, whereas our approach allows creation of much smaller relations for join results.
For the case of group by and aggregation, we believe our approach gives a simpler mapping.

\section{Performance Study}
\label{sec:perf}

We implemented the techniques described in this paper as an extension to the XData system \cite{xdata:2015}.  In this section we present a performance study of the effectiveness of our approach in catching errors in SQL queries, across queries using various SQL features, and across a variety of types of mutations.


Student queries with errors were used to
test earlier versions of XData (e.g. \cite{xdata:2019}). The queries used there were mostly simple queries,
and there was limited diversity in mutations.
Existing benchmarks for query equivalence testing such as Leetcode, Calcite, and Literature \cite{verieql2024arxiv} consist of query pairs where 
most of the pairs are equivalent, and thus cannot 
be used to check for effectiveness in killing of mutants.

We therefore created a collection of more complex queries, along with a large number of non-equivalent mutations, which we call the new XData benchmark [XDataBM].  These queries are based on the University schema \cite{db-book}.


The XDataBM benchmark is based on  
84 original queries, encompassing single-level and multi-level nested queries with and without correlation
among other features.  The distribution of query types is summarized in Table~\ref{table:xdata comp}.
The data generation algorithm produces multiple datasets for each original query, with the initial dataset designed to yield non-empty results.

The tests were run on a computer with
an AMD Ryzen 7 5700G with Radeon Graphics 3.8 GHz, and 16 GB of memory, running Ubuntu 22.04.3 LTS.
As mentioned in Section~\ref{sec:resulttable:size} we cap the size of result tables, with the default cap being 16 tuples.
The data generation algorithm was executed for each original query with a 120 sec. timeout for each targeted dataset.


\begin{table}
\caption{XDataBM Original Queries and Results}
\label{table:xdata comp}
\centering
\resizebox{0.48\textwidth}{!}{
\begin{tabular}{|p{3.5cm}|c|c|c|c|c|}
\hline
\textbf{Query type} & \textbf{\#Queries} & \textbf{\#Mutants} & \textbf{Avg} & \textbf{Avg} & \textbf{Avg Time/} \\
& & & \textbf{\#Datasets} & \textbf{\#Tuples} & \textbf{query(s)}\\
\hline
Single level queries        & 31 &  162  & 8  &  13 & 13.16 \\ \hline
Set operator queries                & 6  &  41 & 6  &  23 & 6.40 \\ \hline
From clause subqueries      & 13 &  48  & 6  &  16 & 7.29\\ \hline
Where clause subqueries     & 34 &  156  & 14 &  25 & 32.62\\ \hline
\textbf{Total}              & 84 & 407 & 10 & 21 & 19.64\\ \hline
\end{tabular}
}

\end{table}

As can be seen from Table \ref{table:xdata comp}, across all query types, the average number of datasets per query is 10, and the average number of tuples per dataset is 21.  The count of number of tuples includes all tuples in input table and result table arrays. 
The average time taken for generating all the datasets for a query is 19.64 sec. 


\subsection{Effectiveness in Killing Mutations}



\begin{table}
\caption{Effectiveness at Killing Mutations}
\label{tab:xdatabm:mutations}
\centering
\resizebox{0.48\textwidth}{!}{
\begin{tabular}{|p{3.5cm}|c|c|c|c|c|}
\hline
\textbf{Mutation type} & \textbf{Total} & \multicolumn{4}{c|}{\textbf{\# of Mutants Killed}} \\ 
\cline{3-6}
& \textbf{Mutants} & \textbf{VEQL} & \textbf{XDataO} & \textbf{USSm} & \textbf{XDataN}  \\
\hline
Selection (comparison) &                   54 & 38  & 31& 40  & \textbf{52}\\ \hline
Join type (Inner vs. Outer) and Join Condition &   73 & 36  & 49&  63 & \textbf{68} \\ \hline
String selection (string comparison) &     78 & 22  & 38&  51 & \textbf{76}\\ \hline
Extra/missing groupby attribute &          18 & \textbf{18}& 12&  14 & \textbf{18}  \\ \hline
Column replacement &                       4 &  2 & 3& \textbf{4}  &  3\\ \hline
Constrained Aggregation &                  10 & 7  & 7& 9 & \textbf{10}\\ \hline
Distinct &                                 7 & \textbf{7}  & 5& \textbf{7}  & 6\\ \hline
Aggregation function &                     35  & 32  & 17& 28  & \textbf{33}  \\ \hline
Missing subquery &                        33  & 12& 20&  25 & \textbf{31}\\ \hline
Subquery connective &                     50  & 11& 35& 40&  \textbf{49} \\ \hline
Set Operator &                            41  & \textbf{41}& 24& 35  &  37 \\ \hline
Null conditions &                         4  & \textbf{4}& 3& \textbf{4}& \textbf{4}\\ \hline
\textbf{Total} &                        407 & 230& 244& 320&  \textbf{378}\\ \hline

\end{tabular}
 }
\end{table}

Each of the 84 queries has associated with it
a number of mutations of different types, with a
total of 407 non-equivalent mutations across the 84 queries.  Table \ref{tab:xdatabm:mutations} shows the
distribution of mutation types; the table also shows the number of these mutations killed by different approaches.
Note that a mutation is killed if the original query and the mutant generate different results on at least one of the datasets.

To study the effectiveness of our approach (XDataN), we compare it with three alternatives.
The first is using the sample University database, which we call USSm, a small database that was manually created by the authors of \cite{db-book}.
The second is the original XData implementation from \cite{xdata:2015}, which we call XDataO.
The last is using the  state-of-the-art query equivalence tool VeriEQL \cite{verieql2024arxiv}
which we call VEQL.  Note however that VeriEQL 
can only be used to compare pairs of queries.  For our target applications, where there is only one given query, VeriEQL is not applicable.\footnote{XData-BM includes queries with natural join, which is not supported by VeriEQL; we therefore rewrote such queries using the inner join operator.}

Overall, XDataN kills around 93\% of the mutations, significantly outperforming USSm, XDataO, and VEQL. 
The reasons for XDataN failing to kill some mutations include lack of support for some mutations in our implementation, and 
timeouts by the solver, indicating the constraints were too complex for it to solve within the timeout period.
Handling a larger class of queries/mutations, and optimizing the constraints to reduce solver time are areas of ongoing work.


It is worth noting that although XDataO outperforms the small university dataset (USSm) for queries presented in \cite{xdata:2015}, XData-BM contains many complex queries with features that XDataO is not able to handle, 
(for example 204 query mutations involve subqueries with correlation and aggregation), and thus USSm beats XDataO on the queries in XDataBM.  In contrast, XDataN significantly outperforms USSm on these queries.  Further, the queries in XDataBM use constants that are present in USSm; for queries that use
different constants in selection conditions, USSm would perform much worse.

\begin{table}
\caption{Mutant Killing by Subquery Type}
\label{tab-subq-types}
\centering
\resizebox{0.4\textwidth}{!}{
\begin{tabular}{|l|c|c|c|}
\hline
\textbf{Query Type}  & \textbf{Total} & \multicolumn{2}{c|}{\textbf{Killed Mutants}}  \\ 
\cline{3-4}
&  \textbf{Mutants} & \textbf{VEQL} & \textbf{XDataN}  \\

\hline
From clause subquery & 48  & 44& \textbf{46} \\
\hline
Where clause with correlation & 80  & 0 & \textbf{79} \\
\hline
Where clause without correlation & 33  & 0 & \textbf{31}\\
\hline
Scalar/In/Not In Subqueries & 43& \textbf{42}& 36\\ 
\hline
Total & 204& 86& \textbf{192}  \\
\hline

\end{tabular}
}
\end{table}

Table~\ref{tab-subq-types} shows the effectiveness of XDataN and VeriEQL on killing mutations of different types of subqueries.   VeriEQL is unable to handle WHERE clause subqueries with EXISTS connective, with or without correlation, whereas XDataN could kill almost all mutations of these and other types of subqueries.

We also ran experiments on the 22 queries from the TPC-H benchmark, with 213 mutants. XDataN is currently able to handle 15 out of 22 queries, and kills 175 out of 213 mutants.  We are currently working on extending XDataN to support the remaining queries, and mutation types that
are currently not supported.



\subsection{Execution Time}


The average time taken by XDataN to generate all datasets for a query, across all the queries we considered, was 19.64 sec., 
while the average time to generate each dataset is 2.04 sec. 

The datasets generated by XDataN can kill a large class of mutations, whereas VeriEQL is executed for pairs of queries, and generates a dataset showing non-equivalence for that pair.
Thus, the goals are different, and a direct comparison of execution time between XDataN and VeriEQL is not meaningful.
However, to get some idea of the relative performance on creating and solving constraints, the average time taken by VeriEQL to check for equivalence for a pair of queries is 3.51 seconds, can be compared to the 2.04 seconds taken on average by XDataN to generate each dataset.

To study the impact of increasing query size on the cost of dataset generation, we created queries which joins the instructor relation with an increasing number of instances of the advisor relation, with join conditions equating the \textit{advisor.s\_id} values of all copies of the relation, and for one copy of the advisor relation, \textit{advisor.i\_id} is equated to \textit{instructor.id} which is a foreign key join.  Also since the number of tuples in result tables is more for cross products than for joins with join conditions, we also created a version of these queries with the same relations, but without any join conditions.

\begin{table}
\centering
\caption{Execution Time}
\label{table:join:time}
\resizebox{.3\textwidth}{!}{
\begin{tabular}{|c|c|c|c|c|}
\hline
\textbf{\#Tables} & \multicolumn{2}{c|}{\textbf{Avg. Time/DS (s)}}  & \multicolumn{2}{c|}{\textbf{Avg. Tuples/DS}}\\

\cline{2-5}
 & \textbf{~~~~Join~~~~} & \textbf{Cross} & \textbf{~~~~Join~~~~} & \textbf{Cross} \\
\hline

4 & 0.34 & 0.48 & 17 & 23 \\
\hline
5 & 0.56 & 0.73 & 21 & 26 \\
\hline
6 & 0.58 & 1.31 & 23 & 35 \\
\hline
7 & 0.60 & 1.70 & 24 & 36 \\
\hline
\end{tabular}
}
\end{table}

Table~\ref{table:join:time} shows the average time taken per dataset and the average number of tuples per dataset, where tuples for a dataset is the sum of the input table and result table array sizes. 
Results for the join and cross product cases are shown separately.  The results show that performance is good even with increasing number of joins.

We note that the forward mapping and backward mapping constraints of the result table described in Section~\ref{sec:join:result} employ nested quantifiers with forall and exist. 
Past experience with XData showed that when quantifying over a fixed size array, unfolding of constraints often results in considerable speedup; for example instead of
asserting ``for all tuples $t$ in $r$, predicate $P(t)$ is true'', if $r$ has 4 tuples we instead assert ``$P(r[1]) \wedge P(r[2]) \wedge P(r[3]) \wedge P(r[4])$''.   Unfolding of exists quantifiers is similarly done by using or (``$\vee$'') instead of
and ($\wedge$).  All the numbers we present are with unfolding.  We also measured the performance of constraint solving without unfolding, and found it to be significantly worse, with execution timing out on many cases; for brevity we omit details.

\section{Conclusion and Future Work}

We have described techniques for generation of test data to catch errors in complex SQL queries.  Our performance study shows that our approach is able to catch a wide variety of errors, and outperforms earlier approaches.

Future work includes (i) using data generation for testing output of text to SQL systems, (ii) optimizing the generated constraints to handle larger queries and larger datasets, (iii) extensions to handle more SQL features such as windowing functions, (iv) extensions to handle further classes of mutations, and (v) extending our system to check query equivalence.

\noindent \textbf{Acknowledgements}:
We thank the all the students who have contributed to the implementation of XData since the version described in \cite{xdata:2015}, who include:
Bikash Chandra, 
Saurabh Chaturvedi,  
Ravi Shankar,  
Sai Balaji Polakampalli, 
Rahul Sharma, 
Pooja Gayakwad, 
Deeksha Kasture, 
and Kumaran Kartikeyan,  





\clearpage

\appendix

\section{Implementation}

In this section, we describe the implementation details of data generation using the query result table approach. The data generation algorithm is described in Section \ref{sec:dataset:gen}. We provide further details in this section.  In particular we provide an overview of each of the key functions in the implementation, to provide links between the high level concepts described in this paper and the actual code base.

\subsection{Pre-processing}
A query is parsed into SQL query tree structure using jsqlparser-5.0.
A unique identification number is assigned to each query block, ensuring that each result table gets a unique name.


\eat{
The following is the list of the main functions for parsing the SQL query. All these functions help create a data structure called \textbf{queryStructure}.
\begin{itemize}
    \item buildQueryStructure  
\item buildQueryStructureJSQL 
\item ProcessSelect 
\item processFromClause 
\item processWhereClause 
\item flattenAndSeparateAllConds 
\item processProjectionList 
\item processGroupByList 
\item processHavingClause 
\item processOrderByList 
\end{itemize}
}

\subsection{Framework for mutation enumeration}

Function \textsc{generateDatasetsToKillMutations} acts as a driver function to enumerate over mutation types and generate corresponding datasets. The driver function uses the function \textsc{generate}\-\textsc{DatasetForNonEmptyDataset} for non-empty data generation, and the function \textsc{generateCons\-traintsToKillMutations} for data generation to kill mutants respectively. In the next section, we describe details of non-empty data generation, which is extended for mutation killing.

\subsection{Non-empty data generation}

Function \textsc{generateDatasetForNonEmptyDataset} uses the function \textsc{generateDataForOriginalQuery} as the main driving function.  In the rest of this section we discuss the major components of non-empty data generation and how they are implemented.

\subsubsection{\textbf{Header generation}} 

Header includes enumerated datatype declaration considering equivalence classes, defining check constraints for the columns, and custom-defined comparison functions, Null values, and null check functions are defined here. Function \textsc{getTupleTypesForSolver} gets the definition of all the arrays of tuples i.e base relations. This definition is created by using Z3 API (using Z3 context defined in ConstraintGenerator class). Function \textsc{generateSolver\_Header} function is used to create the header which includes everything mentioned above.

\subsubsection{\textbf{Database and null constraints}}

Function \textsc{generateNullandDBConstraints} emulates database constraints in Z3 including Foreign keys, primary keys, and domain constraints for columns (including check constraints, is null constraints, etc).

\subsubsection{\textbf{Result table}}

For this part, we first need some preprocessing done to handle correlated nested subqueries, aliasing in from clause subqueries, etc.

First, let's see what we need to process within FROM clause subqueries. So a projected column(nodes) on an outer level can be coming from a FROM clause subquery, and another column may be coming from a base input table. To differentiate between such columns, we maintain an attribute \textsc{fromBelowLevelFClauseSQ} in the Node type. This is specifically worked out for columns in the projection clause. Function \textsc{ProcessProjectedColumnsFromFCla\-useSQ} is a recursive function that does this task.

Now we discuss the decorrelation for nested where clause subqueries. Function \textsc{TraverseNestedQueryStrcuture} is a driving function that traverses the querystructure recursively. Here we first create a hashmap \textsc{aliasMappingToLevels} which maps each unique table name/table alias to the queryblock level it belongs to. Which helps in identifying correlated conditions. 

The next step is to build correlation details, which iterates over every where clause condition to first find out if the condition is correlated. If it is correlated, function \textsc{isCorrelated} sets the flag "isCorrelated" of Node. We create correlationStrcuture for each correlated node with additional information. All the correlationStructure are then put in a hashmap "correlationHashMap" which is later used to iterate over the correlated conditions. Function \textsc{setLe\-velFor\-ProcessingCorrelationCondition} decides the level at which each correlation condition is added after decorrelation.

Now we look at generating constraints for the result tables. Function \textsc{QueryBlockDetails.TraverseNestedQueryBlock} is a driver function that recursively generates constraints for every query block in a bottom-up fashion (deepest level first). Also puts result tables in "aliasMappingToLevels" with levelStructure.

Function \textsc{segregateSelectionConditionsForQueryBlock}  segregates the selection and string condition as this is important to handle string conditions.

Function \textsc{getConstraintsForQueryBlock} creates JRT, ART, and DRT for each block of query. Section \ref{sec:query-result-table} describes the creation of JRT and ART in depth. In section \ref{par-projection-operation}, we described the projection operation of two types. If a distinct clause is present in projection, we create a distinct result table for all types of query blocks to remove the duplicates in projected columns. If the projection is without distinct, we create a projection result mainly for from clause subqueries,

For each block, a join operation result table is first created. Aggregate results and projection results are created subsequently. 
Fu\-nction \textsc{getConstraintsForJoinsInSameQueryBlockWithCo\-unt} 
 creates a JRT table with the help of  function \textsc{getForwardPass\-JoinandSelectionConstraints}, and function
\textsc{getBackward\-Pass\-Join\-andSelectionConstraints}. 

Function \textsc{GetGroupByConstraintsForSubqueryTableWithCount} is used to generate ART constraints with the help of functions \textsc{getPrimaryKeyConstraintsForGSQTable}, 
\textsc{generateAggregateConstraints}, \textsc{generateAggregateConstraints}, \textsc{generateForwardPassConstraints}, and \textsc{generateBackwardPassConstraints}. Later it deals with having clause constraints and writing corresponding selection conditions on ART. To handle the count(*) aggregate function, the aggregate function now contains a list of all the columns.
Function \textsc{GetDistinctClauseConstraintsForSubqueryTableWithCount} creates DRT (with and without a distinct clause in projection.)

Function \textsc{cutRequiredOutputForSMTWithAPI} is a function that executes the SMT constraints generated using Z3 API to create a model if the constraints are satisfiable. Later the same functions parse the model using the API to get .SQL file containing insert statements for generated data.

\subsection{Data generation to kill mutants}
Function \textsc{generateConstraintsToKillMutations} is called for each mutation type separately, first collecting all the target locations and mutation of the target location for the corresponding mutation type. For each of the collected data points (i.e., a mutant) we generate a dataset to kill the mutation. Functions \textsc{traverseQueryBlocksToCollectTargetMutantNodes} and \textsc{generateDatasetForSpecificMutation} are the helping functions to perform the above tasks. For each mutation type, Function \textsc{generateDataForQueryMutants} is used to generate data with little preprocessing. The same Z3 constraints are used as discussed for non-empty data generation with minor changes for each mutation.
\section{Sample Constraints}
In this section, we give samples of currently implemented Z3 constraints. The constraints generated are for the following query. 
\begin{verbatim}
    SELECT COUNT(ID) 
    FROM STUDENT JOIN DEPARTMENT 
          ON (STUDENT.DEPT_NAME=DEPARTMENT.DEPT_NAME) 
    GROUP BY STUDENT.DEPT_NAME ;
\end{verbatim}

\subsection{Declaring Datatypes}

We show below how our current implementation declares the datatype of string valued attribute dept\_name, along with sample constants.  Currently, constants have relation names prefixed to make them unique; we plan to remove the relation names and update the implementation to match the description in Section \ref{data-type}. 
Note that the null value is explicitly declared.

\begin{verbatim}
(declare-datatypes ((DEPT_NAME 0)) 
    (((_DEPT_uNAME__Finance) (_DEPT_uNAME__History) 
    (_DEPT_uNAME__Physics) (_DEPT_uNAME__Music) 
    (_DEPT_uNAME__Biology) (NULL_DEPT_NAME_1))))
\end{verbatim}  

\subsection{Is Null Chcck}

The following function helps check if a value for dept\_name is null.
\begin{verbatim}
(declare-fun ISNULL_DEPT_NAME (DEPT_NAME) Bool)
    (assert (forall ((dept_name DEPT_NAME))
    (= (ISNULL_DEPT_NAME dept_name) 
        (or (= dept_name NULL_DEPT_NAME_1)))))
\end{verbatim}

\subsection{Non Null Constraints}

Not null constraints on an attribute of a relation are enforced as follows, using the is-null check function.
\begin{verbatim}
(assert (forall ((ipk0 Int)) 
    (and (not ( ISNULL_DEPT_NAME (department_DEPT_NAME0 
    (select O_department ipk0)))))))
\end{verbatim}

\subsection{Comparison Operation for Enumerated Datatypes}

In Section \ref{string-type}, we sketched how to create a mapping function of enumerated values to integers, to enable comparison for enumerated datatype.   The following functions implement the mapping and comparison for dept\_name.

\begin{verbatim}
(declare-fun DEPT_NAMEMap ((DEPT_NAME)) Int)
    (assert (= ( DEPT_NAMEMap _DEPT_uNAME__Biology ) 1 ))
    (assert (= ( DEPT_NAMEMap _DEPT_uNAME__Finance ) 2 ))
    (assert (= ( DEPT_NAMEMap _DEPT_uNAME__History ) 3 ))
   (assert (= ( DEPT_NAMEMap _DEPT_uNAME__Music ) 4 ))
    (assert (= ( DEPT_NAMEMap _DEPT_uNAME__Physics ) 5 ))

(declare-fun gtDEPT_NAME ((DEPT_NAME) (DEPT_NAME)) Bool)
(assert (forall  ((x DEPT_NAME) (y DEPT_NAME)) 
(= (gtDEPT_NAME x y) (> (DEPT_NAMEMap x)(DEPT_NAMEMap y))
)))

(declare-fun ltDEPT_NAME ((DEPT_NAME) (DEPT_NAME)) Bool)
(assert (forall  ((x DEPT_NAME) (y DEPT_NAME)) 
(= (ltDEPT_NAME x y) (< (DEPT_NAMEMap x)(DEPT_NAMEMap y))
)))
\end{verbatim}

\subsection{Domain and Check Constraints}

In the schema declaration, attribute TOT\_CRED has a check constraint to ensure the value is between 0 and 1000.
The check constraint is enforced by first declaring a function that checks a single value, and then applying the function to all tuples.  Here -99999 represents the null value for the integer data type, which may be changed to minimum int val in future.

\begin{verbatim}
(declare-fun checkTOT_CRED (Int) Bool)
(assert (forall ((i_TOT_CRED Int))
    (= (checkTOT_CRED i_TOT_CRED)
        (or (and (> i_TOT_CRED (- 1)) (< i_TOT_CRED 1001))
         (= i_TOT_CRED (- 99999))))))
\end{verbatim}

The following constraints are imposed on corresponding columns using check functions defined as above.

\begin{verbatim}
(assert (forall ((i Int))
    (let ((a!1 (or 
    (checkTOT_CRED (student_TOT_CRED3 
    (select O_student i)))
    (ISNULL_TOT_CRED (student_TOT_CRED3 
    (select O_student i))))))
(=> (and (<= 1 i) (<= i 1)) (and a!1)))))
\end{verbatim}

\subsection{Table Declaration}
\label{table-decl-const}

The current implementation for table declaration is as follows, which can be optimized further as described in Section \ref{data-type}.
The following constraints first create a tuple type containing all the attributes of the relation. In addition to that, we define an array of tuples that corresponds to a relation department.
\begin{verbatim}
(declare-datatypes ((department_TupleType 0)) 
    (((department_TupleType 
        (department_DEPT_NAME0 DEPT_NAME) 
        (department_BUILDING1 BUILDING) 
        (department_BUDGET2 Real) 
        (department_CNT3 Int)))))
    
(declare-fun O_department () 
    (Array Int department_TupleType))

\end{verbatim}
Note that an extra \verb|CNT| field has been added to each tuple, with an extra number for disambiguation; in our actual implementation we use \verb|XDATA_CNT| to avoid the risk of name clashes with existing attributes.

\subsection{Primary Key Constraints}
Primary key constraints for the table department are shown below. The following constraints ensure that for a pair of tuples in the table, either the primary key attribute values are distinct or one of the tuples is invalid. Additionally, we ensure that the \verb|CNT| value of each tuple is either 0 or 1 i.e the tuple is either invalid or unique.
\begin{verbatim}
(assert
(or 
  (not(= 
    (department_DEPT_NAME0 (select O_department 1))
    (department_DEPT_NAME0 (select O_department 2))
  ))
  (= (department_CNT3 (select O_department 1)) 0)
  (= (department_CNT3 (select O_department 2)) 0)
))
  
(assert 
(or 
  (= (department_CNT3 (select O_department 1)) 0)
  (= (department_CNT3 (select O_department 1)) 1)
)
(or
  (= (department_CNT3 (select O_department 2)) 0)
  (= (department_CNT3 (select O_department 2)) 1)
))
     
\end{verbatim}

\subsection{Foreign Key Constraints}

Following are the foreign key constraints where student.dept\_name references department.dept\_name.   Conceptually the constraint needs to enforce that for all students there exists a matching tuple in the department.  For efficiency, the for all and exists quantifiers have been unrolled as follows. 
The forall constraint is unrolled by creating a separate constraint for each student.  The exists constraint is unrolled by requiring that either the first tuple in the department matches, or the second tuple in the department matches, and so on. Here constraints for the first tuple in the student table are shown and the department table contains two tuples whose constraints are defined and named as \verb|a!2| and \verb|a!4| using the let feature of Z3.  Then the foreign key constraint ensures one of the following is true: \verb|a!2| or \verb|a!4| or the foreign key attribute value is null.

\begin{verbatim}
(assert 
(let a!2 
  (and 
    (or 
    (= (student_DEPT_NAME2 (select O_student 1))
    (department_DEPT_NAME0 (select O_department 1)))
    )
    (> (department_CNT3 (select O_department 1)) 0))   
)
(let a!4 
  (and 
    (or 
    (= (student_DEPT_NAME2 (select O_student 1))
    (department_DEPT_NAME0 (select O_department 2)))
    )
    (> (department_CNT3 (select O_department 2)) 0))
)
(or 
   a!2 
   a!4 
   (ISNULL_DEPT_NAME (student_DEPT_NAME2 
        (select O_student 1)))
   (= (student_CNT4 (select O_student 1)) 0))))

\end{verbatim}

\subsection{Join Result Table}

In this section, we give constraints for the join result table. 

\subsubsection{\textbf{Table declaration}}
The table declaration constraints for JRT are the same as mentioned in section \ref{table-decl-const}. All the attributes from input tables are included in JRT with unique names as follows:
\begin{verbatim}
(declare-datatypes ((JRT0_TupleType 0)) 
    (((JRT0_TupleType (JRT0_student__ID0 ID) 
        (JRT0_student__NAME1 NAME) 
        (JRT0_student__DEPT_NAME2 DEPT_NAME) 
        (JRT0_student__TOT_CRED3 Int) 
        (JRT0_student__CNT4 Int) 
        (JRT0_department__DEPT_NAME0 DEPT_NAME) 
        (JRT0_department__BUILDING1 BUILDING) 
        (JRT0_department__BUDGET2 Real) 
        (JRT0_department__CNT3 Int) 
        (JRT0__CNT Int)))))

(declare-fun O_JRT0 () (Array Int JRT0_TupleType))
\end{verbatim}

\subsubsection{\textbf{Input to result table mapping}}

We show below a helper function that allows us to map (equate) attributes from specific tuples of input tables student and departmemnt to specific result table tuples, identified by array indices (\verb|x!0| and \verb|x!1| below).  We subsequently use these helper functions to define the join result table constraints.

\begin{verbatim}
(define-fun JRT0_map_student((x!0 Int) (x!1 Int)) Bool
(and 
    (= (student_ID0 (select O_student x!0)) 
    (JRT0_student__ID0 (select O_JRT0 x!1)))
    (= (student_NAME1 (select O_student x!0))
    (JRT0_student__NAME1 (select O_JRT0 x!1)))
    (= (student_DEPT_NAME2 (select O_student x!0))
    (JRT0_student__DEPT_NAME2 (select O_JRT0 x!1)))
    (= (student_TOT_CRED3 (select O_student x!0))
    (JRT0_student__TOT_CRED3 (select O_JRT0 x!1)))
    (= (student_CNT4 (select O_student x!0))
    (JRT0_student__CNT4 (select O_JRT0 x!1))))
)

(define-fun JRT0_map_department((x!0 Int) (x!1 Int)) Bool
(and 
(= (department_DEPT_NAME0 (select O_department x!0))
   (JRT0_department__DEPT_NAME0 (select O_JRT0 x!1)))
(= (department_BUILDING1 (select O_department x!0))
   (JRT0_department__BUILDING1 (select O_JRT0 x!1)))
(= (department_BUDGET2 (select O_department x!0))
   (JRT0_department__BUDGET2 (select O_JRT0 x!1)))
(= (department_CNT3 (select O_department x!0))
   (JRT0_department__CNT3 (select O_JRT0 x!1)))
)
)
\end{verbatim}

\subsubsection{\textbf{CNT Calculation}}
The multiplicity XDATA\_CNT for a tuple in JRT is either 0 if join conditions are not satisfied, or calculated by multiplying the XDATA\_CNT values of the input tuples.
The first constraint ensures the count values are non-negative, and the second constraint checks that either the count is 0 (invalid tuple) or is equal to the product of the input tuple counts. 

\begin{verbatim}
(assert (forall ((i1 Int)) 
(and 
(>= (JRT0__CNT (select O_JRT0 i1)) 0) 
(or 
    (= (JRT0__CNT (select O_JRT0 i1))
    (* (JRT0_student__CNT4 (select O_JRT0 i1))
        (JRT0_department__CNT3 (select O_JRT0 i1)))) 
    (= (JRT0__CNT (select O_JRT0 i1)) 0))
)))
\end{verbatim}

\subsubsection{\textbf{Blocking duplicates}}

As the duplicate count for each tuple is maintained separately, we avoid creating two duplicate valid tuples by asserting the following constraints that ensure that if two tuples have the same attribute values, then at least one of them is invalid.
\begin{verbatim}
(assert (=> 
    (and (= (JRT0_student__ID0 (select O_JRT0 1))
        (JRT0_student__ID0 (select O_JRT0 2)))
     (= (JRT0_department__DEPT_NAME0 (select O_JRT0 1))
        (JRT0_department__DEPT_NAME0 (select O_JRT0 2)))
     ))    
(or (= (JRT0__CNT (select O_JRT0 1)) 0)
    (= (JRT0__CNT (select O_JRT0 2)) 0)))
\end{verbatim}

\subsubsection{\textbf{Ensuring non-empty result table}}
To ensure our result table is non-empty we assert that at least one tuple in the result table is valid at the outermost level. The following shows how to do it for the case where the table has 2 tuples selected by \verb|(select O_JRT0 1)| and \verb|(select O_JRT0 2)|.

\begin{verbatim}
  (or (> (JRT0__CNT (select O_JRT0 1)) 0)	
      (> (JRT0__CNT (select O_JRT0 2)) 0)	)
\end{verbatim}

\subsubsection{\textbf{Forward mappng}}
In Section \ref{normal-join}, we described the forward mapping function to map tuples from the input table to the result table. For each combination of valid tuples from input tables, we check if the join and selection conditions are true, if yes then it should map to one of the tuples in JRT. If the student has 1 tuple and the department has 2, we can have combinations of tuples in the cross product as (1, 1),(1, 2). The following set of constraints is an example of mapping the student and department table to the result table for combination (1, 1). 
\begin{verbatim}
(assert	 
(=> 
 (and 
   (and 	 
   (= (student_DEPT_NAME2 (select O_student 1)) 
   (department_DEPT_NAME0 (select O_department 1))) 
   (not (ISNULL_DEPT_NAME 
   (student_DEPT_NAME2 (select O_student 1))))
   (not (ISNULL_DEPT_NAME 
   (department_DEPT_NAME0 (select O_department 1)))))
   (> (student_CNT4 (select O_student 1)) 0)
   (> (department_CNT3 (select O_department 1)) 0))
 (or
    (and 
      (> (JRT0__CNT (select O_JRT0 1)) 0)
      (JRT0_map_student 1 1)
      (JRT0_map_department 1 1)
      ))
    (and 
      (> (JRT0__CNT (select O_JRT0 1)) 0)
      (JRT0_map_student 1 2)
      (JRT0_map_department 1 2)
      ))
)
\end{verbatim}
Similarly, constraints are added for other combinations.

\subsubsection{\textbf{Backward Mapping}}
Similar to forward mapping, backward mapping constraints map result table tuples to input tuples as follows.  The constraints ensure that every valid tuple in the result table must satisfy join condition, and further must be mapped to some tuple from each input table.
\begin{verbatim}
(assert (forall ((i1 Int)) 
(=> (> (JRT0__CNT (select O_JRT0 i1)) 0)
(and 
    (> i1 0)
    (<= i1 2)
    (and 
      (= (JRT0_student__DEPT_NAME2 (select O_JRT0 i1))  
        (JRT0_department__DEPT_NAME0 (select O_JRT0 i1))) 
      (not (ISNULL_DEPT_NAME (JRT0_student__DEPT_NAME2
        (select O_JRT0 i1))))
      (not (ISNULL_DEPT_NAME 
         (JRT0_department__DEPT_NAME0 select O_JRT0 i1)
         ))
      )   
    (or (JRT0_map_student 	1	i1))
    (or (JRT0_map_department 1	i1)
        (JRT0_map_department 2	i1))
))))
\end{verbatim}

\subsection{Aggregate Result Table}

In this section, we provide sample constraints for the aggregation result table (ART). 

\subsubsection{\textbf{Table declaration}}
The ART attributes are the group by attributes and aggregated output for each aggregate function.  In this case an aggregated result attribute corresponding to COUNT(ID) is present.

\begin{verbatim}
(declare-datatypes ((ART0_TupleType 0)) 
(((ART0_TupleType (ART0_student__DEPT_NAME DEPT_NAME) 
(ART0_COUNTstudent__ID Int) (ART0_CNT Int)))))

(declare-fun O_ART0 () (Array Int ART0_TupleType))
\end{verbatim}

\subsubsection{\textbf{Primary key constraints}}
In ART, each valid tuple corresponding to a group must be unique.   The following constraints are asserted to ensure this property for every pair of tuples, enforcing that if two tuples are equal on the group by attributes at least one of them should be invalid. 
\begin{verbatim}
(assert 
(=> 
    (and 
        (= (ART0_student__DEPT_NAME (select O_ART0 1))
           (ART0_student__DEPT_NAME (select O_ART0 2)))
    )
    (or 
        (= (ART0_CNT (select O_ART0 1)) 0)
        (= (ART0_CNT (select O_ART0 2)) 0)
    )))
\end{verbatim}
Additionally, the following constraints restrict CNT value of tuples in ART to be either 0 or 1 as follows:
\begin{verbatim}
(assert (forall ((i1 Int)) 
    (or (= (ART0_CNT (select O_ART0 i1)) 0)
    (= (ART0_CNT (select O_ART0 i1)) 1))))
\end{verbatim}

\subsubsection{\textbf{Forward mapping}}
Forward mapping constraints map input tuples to the corresponding groups in ART as follows. 
\begin{verbatim}
(define-fun SQTABLE_0_FORWARD((i1 Int)) Bool
(and 
(=> (> (JRT0__CNT (select O_JRT0 i1)) 0)
(or
(and 
    (= (ART0_student__DEPT_NAME (select O_ART0 1))
     (JRT0_student__DEPT_NAME2 (select O_JRT0 i1)))
     (= (ART0_CNT (select O_ART0 1)) 1))
(and 
    (= (ART0_student__DEPT_NAME (select O_ART0 2))
     (JRT0_student__DEPT_NAME2 (select O_JRT0 i1)))
     (= (ART0_CNT (select O_ART0 1)) 2))
))))
(assert (SQTABLE_0_FORWARD 1))
(assert (SQTABLE_0_FORWARD 2))
\end{verbatim}

Forward mapping constraints are asserted for every tuple in the input table. The helper function constraints ensure that for a particular input tuple, if it is then it must map to one of the valid tuples (i.e. groups) in ART.  The helper function is then asserted for each of the input tuples.

\subsubsection{\textbf{Backward mapping}}

These constraints are used to ensure that for each valid group present in the result table there is a valid tuple in the input table.
\begin{verbatim}
(define-fun SQTABLE_0_BACKWARD((k1 Int))Bool

(=> (and (= (ART0_CNT (select O_ART0 k1)) 1)) 
(or 
    (and (= (ART0_student__DEPT_NAME (select O_ART0 k1))
          (JRT0_student__DEPT_NAME2 (select O_JRT0 1)))
          (> (JRT0__CNT (select O_JRT0 1)) 0))))
          
    (and (= (ART0_student__DEPT_NAME (select O_ART0 k1))
          (JRT0_student__DEPT_NAME2 (select O_JRT0 2)))
          (> (JRT0__CNT (select O_JRT0 2)) 0))))
))) 
(assert (SQTABLE_0_BACKWARD 1))
(assert (SQTABLE_0_BACKWARD 2))
\end{verbatim}
The backward pass constraints are asserted for each tuple in the ART. these constraints ensure that the particular valid tuples in ART correspond to at least one valid tuple in the input table.  In this case the input table JRT0 has 2 tuples so the function ensures that the ART tuple maps to at least one of them, and that tuple is valid.  The helper function is invoked for each of the two tuples in ART.

\subsubsection{\textbf{Aggregate calculations}}
This is the function that calculates COUNT(ID) for each group. Similarly, we create functions to calculate all other aggregate results sum, avg, min, max.
\begin{verbatim}
(define-fun GETAGGVALCOUNTID0((i Int))Bool
  (or 
  (= (ART0_COUNTstudent__ID (select O_ART0 i)) 
    (+ (ite 
        (and 
          (= (ART0_student__DEPT_NAME (select O_ART0 i))
          (JRT0_student__DEPT_NAME2 (select O_JRT0 1)))) 
        (JRT0__CNT (select O_JRT0 1)) 0)
        (ite 
          (and 
            (= (ART0_student__DEPT_NAME (select O_ART0 i))
          (JRT0_student__DEPT_NAME2 (select O_JRT0 2)))) 
          (JRT0__CNT (select O_JRT0 2)) 0)    
    ))
  (= (ART0_CNT (select O_ART0 i)) 0)))
(assert (GETAGGVALCOUNTID0 1))
(assert (GETAGGVALCOUNTID0 2))
\end{verbatim}

The above constraints are asserted for each tuple in ART to calculate COUNT(ID) for each group. 
To calculate the aggregate value for a particular tuple in ART, the helper function \verb|GETAGGVALCOUNTID0| aggregates the counts from all the tuples from the input table (in this case 2 tuples) that match the group, using the if-then-else expression \verb|ite| supported by Z3 to return the count for matching tuples, and 0 for non-matching ones.
The helper function then ensures that if a particular tuple is valid, its aggregate value is computed as above.
Finally, the helper function is asserted for each tuple in ART.

\section{Decorrelation}
\label{app:decorr}

We provide more examples of complex decorrelations
to illustrate how decorrelation works, as also optimizations to make the decorrelated queries simpler.

\subsection{Optimization of Multiple Level Exists with Equality}

We saw earlier in Section~\ref{sec:subquery:where} the following query.

\begin{verbatim}
Q5: SELECT R.A, R.B FROM R                --> Level 0
    WHERE EXISTS (                 
        SELECT S.C FROM S                 --> Level 1
        WHERE S.C = R.C AND EXISTS (
            SELECT * FROM T               --> Level 2
            WHERE T.A = R.A AND T.B = S.B))
\end{verbatim}

We saw that the query can be decorrelated as follows:
\begin{verbatim}
Q5': SELECT R.A, R.B                       --> Level 0
     FROM R 
     SEMI JOIN                           
        (SELECT SQ3.A, SQ3.C
         FROM (SELECT S.B, S.C, R.A        --> Level 1
               FROM S, (SELECT DISTINCT R.A FROM R)
              ) as SQ3
            SEMI JOIN                   
              (SELECT *  FROM T) as SQ2    --> Level 2 
            ON (SQ2.B = SQ3.B AND SQ2.A = SQ3.A)
         ) as SQ1 
     ON (SQ1.C = R.C AND SQ1.A = R.A)
\end{verbatim}

For the case where the correlation variable from two levels above is used in an equality condition, as in \texttt{T.A = R.A}, we can optimize the rewriting by replacing the semijoin with a join.  The optimized
version of the above transformation is as follows:
\begin{verbatim}
Q5":
    SELECT R.A, R.B                       --> Level 0
    FROM R 
    SEMI JOIN (                           --> Level 1
        SELECT S.C AS S_C, SQ2.T_A AS SQ2_T_A  
        FROM S JOIN                       --> Level 2        
             (SELECT DISTINCT T.A as T_A, T.B as T_B
              FROM T ) as SQ2  
            ON (SQ2.T_B = S.B AND SQ2.T_A=SQ1.R_A)
    ) as SQ1 ON (S_C = R.C) AND SQ2_T_A = R.A)
\end{verbatim}
Note that the semijoin of \verb|SQ1| and \verb|SQ2| has been replaced by a join of S and SQ2, to allow the attribute \verb|SQ.T__A| to be available for the semijoin at the outer level.  Further a DISTINCT keyword has been added to the SELECT clause of SQ2 to ensure that the duplicate count of the join result matches that of the original semijoin.

\eat{
\paragraph{\textbf{Not Exists Connective Without Correlation:}}

Consider an example query with a single level where clause subquery with NOT EXISTS connective.

\begin{verbatim}
Q6:
    SELECT R.A, R.B FROM R                 --> Level 0
    WHERE NOT EXISTS (                 
        SELECT S.A  FROM S                 --> Level 1
    );
\end{verbatim}

We decorrelate where clause subquery with NOT EXISTS connective into a from clause subquery using anti semijoin as shown in query Q6'.

\begin{verbatim}
Q6':
    SELECT R.A, R.B                        --> Level 0
    FROM R ANTI SEMI JOIN (             
        SELECT S.A  FROM S                 --> Level 1 
    );
\end{verbatim}
$SQ_1$ represents a result table for subquery level 1 and acts as one of the input tables for $SQ_0$. R ANTI SEMI JOIN $SQ_1$ is performed as discussed in Section~\ref{sec:join:result}.
}

Now consider the case of multi-level nested subqueries with non-immediate correlation i.e between level 2 and level 0.

\begin{verbatim}
Q10:
    SELECT R.A, R.B FROM R                --> Level 0
    WHERE EXISTS (                 
        SELECT S.C FROM S                 --> Level 1
        WHERE NOT EXISTS (
            SELECT * FROM T               --> Level 2
            WHERE T.A = R.A AND T.B=S.B
        )
    );
\end{verbatim}

The above query can be decorrelated as below.
Note that the parameter values from the outermost level query are provided to the the Level 2 query by adding a cross product (cross join) with distinct R.A values from R.  

\begin{verbatim}
Q10':
    SELECT R.A, R.B                       --> Level 0
    FROM R 
    SEMI JOIN (                           --> Level 1
        SELECT SQ2.T__A AS SQ2_T__A, R.A
        FROM (S CROSS JOIN (SELECT DISTINCT R.A FROM R))
        ANTI SEMI JOIN (                  --> Level 2            
            SELECT T.A as T__A, T.B as T__B FROM T) as SQ2  
            ON (SQ2.T__B = S.B AND SQ2.T__A=R.A)
    ) as SQ1 ON (SQ1.SQ2_T__A = R.A)
\end{verbatim}

\eat{
The above query can be further simplified as below.
\begin{verbatim}
Q10'':
    SELECT R.A, R.B                       --> Level 0
    FROM R 
    JOIN (                                --> Level 1
        SELECT SQ2.T__A AS SQ2_T__A, R.A
        FROM (S CROSS JOIN R)
        ANTI SEMI JOIN (                  --> Level 2            
            SELECT T.A as T__A, T.B as T__B FROM T) as SQ2  
            ON (SQ2.T__B = S.B AND SQ2.T__A=R.A)
    ) as SQ1 ON (SQ1.SQ2_T__A = R.A)
\end{verbatim}
}

\paragraph{\textbf{Nested Subqueries With Non-Equi Correlation:}}

Consider the following query whose subquery has non-equijoin conditions on correlation variables.

\begin{verbatim}
Q11:
    SELECT R.A, R.B FROM R                --> Level 0
    WHERE EXISTS (                 
        SELECT S.C FROM S                 --> Level 1
        WHERE NOT EXISTS (
            SELECT * FROM T               --> Level 2
            WHERE T.A > R.A AND T.B > S.B
        )
    );
\end{verbatim}

As before, the parameters from the level 1 query have to be passed to the level 1 query by adding a join with R, and used in the anti semijoin condition. 

\begin{verbatim}
Q11':
    SELECT R.A, R.B                       --> Level 0
    FROM R 
    SEMI JOIN (                           --> Level 1
        SELECT SQ1.R_A AS SQ2_T_A
        FROM S, R r1
        ANTI SEMI JOIN (                  --> Level 2    
            SELECT T.A as T_A, T.B as T_B FROM T) as SQ2  
            ON (SQ2.T_B > S.B AND SQ2.T_A > r1.A)
    ) as SQ1 ON (SQ1.SQ2_T_A = R.A)
\end{verbatim}

\paragraph{\textbf{Nested Subqueries With Correlation and Aggregate:}}

Consider the cases where the inner subquery contains aggregation and correlation with the outer query.
\begin{verbatim}
Q12:
    SELECT COUNT(*) FROM R                --> Level 0              
    WHERE NOT EXISTS (
        SELECT SUM(S.A) FROM S            --> Level 1
        WHERE S.B = R.B
        GROUP BY S.A HAVING SUM(S.C) <= 0)
    GROUP BY R.A ;
\end{verbatim}

As we decorrelate the nested subquery and transform it into a from clause subquery using anti semijoin with correlation condition as join condition, SQ1 needs to be grouped by correlation attribute as well. Hence we add the correlation attribute to group by clause in the inner level query.
\begin{verbatim}
Q12':
    SELECT COUNT(*) FROM R                --> Level 0
    ANTI SEMI JOIN (                      --> Level 1
        SELECT SUM(S.A), S.B as S__B FROM S
        GROUP BY S.A, S.B HAVING SUM(S.C) <= 0) as ASQ1
    ON (ASQ1.S_B=R.B) GROUP BY R.A ;
\end{verbatim}


\end{document}